\documentclass[12pt]{article}
\usepackage{color}
\usepackage{axodraw}
\newcommand{\beq}{\begin{equation}}
\newcommand{\eeq}{\end{equation}}
\newcommand{\beqa}{\begin{eqnarray}}
\newcommand{\eeqa}{\end{eqnarray}}
\newcommand{\beqar}{\begin{eqnarray*}}
\newcommand{\eeqar}{\end{eqnarray*}}

\newcommand{\al}{\alpha}
\newcommand{\be}{\beta}

\def\non          {\nonumber}
\def\ha           {\mbox{$\frac{1}{2}$}}

\def\s  {\sigma}

\def\Tr           {\mbox{\rm Tr}\,}
\def\STr          {\mbox{\rm STr}\,}

\def\cd           {{\cdot}}
\def\ran          {\rangle}
\def\lan          {\langle}
\def\fsH    {H\!\!\!\!/\,}
\newcommand{\del}{\delta}

\newcommand{\eps}{\epsilon}
\newcommand{\ga}{\gamma}

\newcommand{\inn}{\!\cdot\!}

\newcommand{\lam}{\lambda}

\newcommand{\z}{\zeta}

\newcommand{\labell}[1]{\label{#1}} 
\newcommand{\reef}[1]{(\ref{#1})}
\newcommand\prt{\partial}
\newcommand\veps{\varepsilon}

\newcommand\cL{{\cal L}}
\newcommand\cD{{\cal D}}

\newcommand\cT{T}

\def\sst#1{{\scriptscriptstyle #1}}
\def\0{{\sst{(0)}}}
\def\1{{\sst{(1)}}}
\def\2{{\sst{(2)}}}
\def\3{{\sst{(3)}}}
\def\4{{\sst{(4)}}}
\def\5{{\sst{(5)}}}
\def\6{{\sst{(6)}}}
\def\7{{\sst{(7)}}}
\def\8{{\sst{(8)}}}

\let\z=\zeta    
\let\s=\sigma   \let\f=\phi

\newcommand{\bea}{\begin{eqnarray}}
\newcommand{\eea}{\end{eqnarray}}
\begin{document}
\baselineskip 16pt%
\begin{titlepage}
\vspace*{1mm}%

\vspace*{3 mm}
\vspace*{5 mm}%
\center{ {\bf {Universality in all-$\alpha'$ order corrections to BPS/non-BPS
\\
brane world volume theories}
}}\vspace*{3mm} \centerline{{\Large {\bf  }}} \vspace*{.1mm}
\begin{center}
{Ehsan Hatefi$^{\,\dagger}$ and I. Y. Park$\,^*$}

\vspace*{0.8cm}{ {\it
International Centre for Theoretical Physics\\
 Strada Costiera 11, Trieste, Italy $^{\dagger}$ \\
ehatefi@ictp.it}}

\vspace*{0.4cm}
{\it Department of Natural and physical Sciences,
Philander Smith College

Little Rock, AR 72223, USA $^*$ \\
inyongpark05@gmail.com } \vspace*{.3cm}
\end{center}
\begin{center}{\bf Abstract}\end{center}
\begin{quote}
Knowledge of all-$\alpha'$ higher derivative corrections to
leading order BPS and non-BPS brane actions
 would serve in future endeavor of determining the complete form of the non-abelian
BPS and tachyonic  effective actions.
In this paper, we note that there is a universality in the all-$\alpha'$ order corrections
to BPS and non-BPS branes. We compute amplitudes between one Ramond-Ramond $C$-field
vertex operator and several SYM gauge/scalar vertex operators. Specifically, we evaluate in closed form string correlators of two-point amplitudes $\cal A^{C\phi}$, $\cal A^{C A}$, a three-point
amplitude $\cal A^{C\phi\phi}$, and a four-point amplitude $\cal A^{C\phi\phi\phi}$.
We carry out pole and contact term analysis.
In particular we reproduce some of the contact terms and the infinite massless poles of $\cal A^{C\phi\phi\phi}$ by SYM
vertices obtained through the universality.

\end{quote}
\end{titlepage}

\section{Introduction}
 D-branes are sources for closed string RR fields in string theory \cite{Polchinski:1995mt}.
 Many results on their properties have been obtained \cite{Polchinski:1996na}.
Without the higher derivative corrections in string theory, the action takes the Born-Infeld
\cite{Tseytlin:1999dj,Myers:1999ps}
and Wess-Zumino (WZ) forms \cite{li1996a}. A recent
comprehensive discussion on the higher derivative corrections
to  BPS and non-BPS effective actions can be found, e.g., in works of \cite
{Hatefi:2010ik} and \cite {Hatefi:2012wj}. An attempt at finding universal higher
derivative corrections was made \cite{Hatefi:2012ve}, and
 higher derivative corrections to WZ couplings for
non-BPS branes were obtained in
\cite{Garousi:2007fk} and \cite{Garousi:2008ge}. Roles played by D-branes were reviewed,
 e.g., in \cite{Polchinski:1996nb} and \cite{Vafa:1997pm}.

\vspace{.2in}

There has been recent progress in string amplitude computation
( see \cite{Hashimoto:1996bf} for an early development) and its matching
with the corresponding low energy DBI computation \cite{Hatefi:2010ik}, where in
the amplitude of one RR closed string vertex and
three open string gauge field vertex operators
has been computed in detail. {A
related analysis for higher derivative corrections of two scalar and
two tachyon vertex operators for a non-BPS case to all orders of $\alpha'$ was carried out
in \cite{Hatefi:2012wj}}. Subsequently,
the amplitude of one RR closed string vertex and two open
string gauge fields and one open string scalar field vertex
operator was computed in \cite{Hatefi:2012ve}. There,
 we found all higher derivative corrections to the two-gauge field and two-scalar
field couplings in a closed form by explicit S-matrix computation.
The all-$\alpha'$ order vertices of SYM were determined by analyzing the poles and contact terms.

\vspace{.2in}

One of the motivations behind our recent and present works is to better understand the full closed form of the non-abelian DBI action.
The closed form of non-abelian DBI action has been evasive despite the continued efforts so far
(see for instance \cite{Koerber:2002zb,Keurentjes:2004tu}).
With an ultimate goal of determining the closed
form of the non-abelian action, we continue collecting "data"; we
analyze another set of amplitudes involving one RR $C$-field vertex
and a few open string vertices. The data will
serve as a useful guide to determining the full non-abelian
DBI action. (Or if the full DBI action is discovered by another
method, the data will provide useful cross checks.)

\vskip 0.1in
In this paper we analyze amplitudes of one RR vertex operator
$C$  and a few scalar field vertex operators in closed form;
   after warming up with $<V_C V_\phi>$ and $<V_C V_\phi V_\phi>$ (we also compute $<V_C V_A>$), we analyze the case of $<V_C V_\f V_\f V_\f>$. Although $<V_C V_\f V_\f V_\f>$ was considered in \cite{Garousi:2000ea}, we have decided to revisit this amplitude for several reasons.
  Firstly, the final form of our amplitude computation is different from that of \cite{Garousi:2000ea}. {Several subtleties are involved in the computation; we will note them based on the recent work
\cite{Hatefi:2010ik}.}
We present all details necessary to derive our results below. In addition, we carry out all infinite poles and some of the contact term
analysis, and determine the corresponding SYM vertices. This part of the analysis was not carried out in \cite{Garousi:2000ea}. {Our computation is based on Wick-like contractions given in \cite{Liu:2001qa,Hatefi:2010ik}}. Secondly, the amplitude is relatively simple yet expected to serve as another test ground for universality, a property that we expect to be a useful asset in future understanding of the closed form of DBI action.
The results of \cite{Hatefi:2010ik,Hatefi:2012wj,Hatefi:2012ve} suggest a method for finding all-$\alpha'$ order corrections to
BPS (and non-BPS) DBI actions.
  The method is based on a universality in the pattern of all-$\alpha'$ order extension of
the vertices in the BPS (and non-BPS) brane actions.\footnote{
One may wonder if it is possible
 to find higher derivative corrections of
BPS branes by
applying T-duality to previously known results. T-duality in open string loop computations is subtle and is not as effective as in tree amplitudes \cite{Park:2008sg} (in particular, see footnote 4 therein).
 Although we are considering tree-level diagrams, the amplitudes share certain attributes
 of open string loop diagrams due to the presence of a closed string.
 Instead of relying on T-duality, we note that there is a persisting pattern in the
 higher order derivative corrections and that one may use the pattern as a prescription for determining
 the forms of the higher order corrections. As we will discuss, the prescription works not only for
 BPS cases but also for non-BPS cases. Recall that the forms of vertex
operators for tachyons are quite different from gauge/scalar field
vertex operators. We will comment on T-duality related issues in the conclusion.
}
   Utilizing the pattern, we determine\footnote{See \cite{Hatefi:2010ik} for an earlier attempt.} closed-form higher derivative
   corrections to four-scalar interaction vertex, and confirm them with an
   explicit Feynman diagram computation.

\vspace{.2in}

  The third reason for considering $<V_C V_\f V_\f V_\f>$ is its relevance for Myers' effect.
We analyzed $<V_C V_A V_A V_\f>$ in \cite{Hatefi:2012ve}, and new WZ couplings were obtained.  The result has been applied in \cite{Hatefi:2012sy} to understand the M5 brane
$N^3$ entropy scaling. In particular, $(31)$ of \cite{Hatefi:2012ve} was discussed in sec 4.2 of \cite{Hatefi:2012sy}. Similarly, some of the new WZ couplings that we find below are expected to play a role in the higher $\alpha'$ contributions to the entropy of various brane configurations.
In fact we have noticed that Myers' terms can be realized in a peculiar manner from the coupling between open and closed strings. They come from the closed string coupling to a lower dimensional branes. The lower dimensional branes can be viewed as soliton solutions of the branes that one started with. We also have obtained solutions that represent dissolution of lower dimensional brane inside of higher dimensional brane which is a manifestation of Myers' terms \cite{Hatefi:2012wz}. For $D(-1)/D3$ system, it could be higher $\alpha'$-order dielectric effect that is responsible for the $N^2$ behavior \cite{Hatefi:2012sy} and \cite{Hatefi:2012 tt}.

\vspace{.1in}

\vspace{.1in} The rest of the paper is organized as follows. In
section 2, we start by analyzing two-point amplitudes $<V_CV_\phi>$ and $<V_CV_A>$.
Then we compute a three-point
amplitude between one RR and two scalar fields $<V_CV_\phi V_\phi>$, and work out
with contact terms. We determine the
corresponding field theory vertices that reproduce {all infinite} contact terms.\footnote{The
 corresponding analysis for one RR, one gauge field and one scalar field
was done in the appendix of \cite{Hatefi:2012ve}.} In
section 3, which contains the main result of this work, we carry out the
analogous analysis for $<V_C V_\f V_\f V_\f>$. First, we compute the string amplitude in closed form
and analyze the infinite poles. Then we review the infinite extension results of \cite{Hatefi:2012wj,Hatefi:2010ik,Hatefi:2012ve} in which various amplitudes
of two tachyons and two scalar fields, four gauge fields, two gauge fields and two scalar fields
were analyzed. We notice a persisting pattern in the form of higher derivative corrections in those results.
   Taking the pattern as a prescription we determine, with correct coefficients, the all-order higher derivative corrections to the SYM
   four-scalar couplings. We further comment on T-duality related aspects in the conclusion after summarizing the results of the
present work.

\section{Analysis of $<V_C V_\phi>$, $<V_C V_A>$ and $<V_C V_\phi V_\phi>$ }

We start by considering relatively simple
amplitudes $<V_C V_\phi>$, $<V_C V_A>$, and $<V_C V_\phi V_\phi>$
as a warm-up, and give more details for $<V_C V_\phi V_\phi>$.
After computing the string amplitudes, we determine through inspection
the SYM vertices that reproduce the momentum expansion
of the string amplitudes. There are patterns in the field theory vertices
that reproduce the leading order poles and contact terms that allow
 one to find the all-order extensions of these vertices. The patterns were present in the previous
 works \cite{Hatefi:2010ik}, \cite{Hatefi:2012wj}, \cite{Hatefi:2012ve} and \cite{Garousi:2008ge} and they persist in the present work. We will see the first example of the pattern in $<V_C V_\phi V_\phi>$ and more examples in section 3.

\subsection{ $<V_CV_\phi>$ }

Here we consider BPS amplitudes in flat space and set all  background fields to zero. The (0)- and (-1)- picture vertex operators for the scalar fields, gauge fields
and the RR $C$-field vertex operator in (-1)-picture are given by\footnote{We keep $\alpha'$
explicitly in this work. We can set $\alpha'=2$ on the string theory side to simplify the computations. Some of our conventions were summarized, e.g., in Appendix A of \cite{Hatefi:2012ve}.}
\beqa
V_{\phi}^{(0)}(x) &=& \xi_{i}\bigg(\partial
X^i(x)+\alpha' ik\cd\psi\psi^i(x)\bigg)e^{\alpha' ik\cd X(x)},
\nonumber\\
V_{\phi}^{(-1)}(y) &=&\xi_j \psi^j(y) e^{-\phi(y)} e^{\alpha' ik\cd X(y)},
\nonumber\\
V_{A}^{(-1)}(x_1) &=&\xi_a\psi^a(x_1) e^{-\phi(x_1)} e^{\alpha'iq\cd X(x_1)}
\nonumber\\
V_{C}^{(-\frac{1}{2},-\frac{1}{2})}(z,\bar{z})&=&(P_{-}\fsH_{(n)}M_p)^{\al\be}e^{-\phi(z)/2}
S_{\al}(z)e^{i\frac{\alpha'}{2}{p}\cd X(z)}e^{-\phi(\bar{z})/2} S_{\be}(\bar{z})
e^{i\frac{\alpha'}{2}{p}\cd D \cd X(\bar{z})},
\label{d4Vs3}
\eeqa
 where $(k,q,p)$ are the momenta
of the scalar field, gauge field and $C$-field. They should satisfy the on-shell condition
$k^2=q^2=p^2=0$.
Our notation is such that
 the spinorial
indices should be raised by the charge conjugation matrix, $C^{\alpha\be}$
\beqa
(P_{-}\fsH_{(n)})^{\al\be} =
C^{\al\del}(P_{-}\fsH_{(n)})_{\del}{}^{\be}
\eeqa
In particular, the traces are defined as the following
\beqa
\Tr(P_{-}\fsH_{(n)}M_p\gamma^{k})&\equiv & (P_{-}\fsH_{(n)}M_p)^{\alpha\beta}(\gamma^{k}C^{-1})_{\alpha\beta}
\nonumber\\
\Tr(P_{-}\fsH_{(n)}M_p\Gamma^{jai})&\equiv &
(P_{-}\fsH_{(n)}M_p)^{\alpha\beta}(\Gamma^{jai}C^{-1})_{\alpha\beta}
\eeqa
where $P_{-}$ is a projection operator and its definition is $P_{-} = \ha (1-\ga^{11})$, and the definition of RR field strength is
\begin{displaymath}
\fsH_{(n)} = \frac{a
_n}{n!}H_{\mu_{1}\ldots\mu_{n}}\ga^{\mu_{1}}\ldots
\ga^{\mu_{n}}
\ ,
\non\end{displaymath}
with $n=2,4$ for type IIA and $n=1,3,5$ for type IIB. $a_n=i$ for IIA and $a_n=1$ for IIB theory.
To work with standard holomorphic world sheet correlators,
we embed the usual doubling trick. To see doubling trick and more details on correlation functions between spin operators and some currents and/or fermion fields we refer to Appendix A of \cite{Hatefi:2012wj}.
 One may use holomorphic correlators for the world-sheet fields $X^{\mu},\psi^{\mu}, \phi$
\begin{eqnarray}
\lan X^{\mu}(z)X^{\nu}(w)\ran & = & -\eta^{\mu\nu}\log(z-w) , \non \\
\lan \psi^{\mu}(z)\psi^{\nu}(w) \ran & = & -\eta^{\mu\nu}(z-w)^{-1} \ ,\non \\
\lan\phi(z)\phi(w)\ran & = & -\log(z-w) \ .
\labell{prop}\end{eqnarray}

Note that in this paper we do not fix the over all signs of the equations.

$\frac{}{}$

Let us consider
\begin{eqnarray}
{\cal A}^{C\phi} & \sim & \int dx_{1}dzd\bar{z}\,
  \lan V_{\phi}^{(-1)}{(x_{1})}
V_{RR}^{(-\frac{1}{2},-\frac{1}{2})}(z,\bar{z})\ran,\labell{cpsstring}\eeqa
where the open string vertex operators ( for this amplitude just $x_1$) should be inserted at the boundary of disk world sheet
and the closed string vertex operator such that $z=x+iy$, $\bar z=x-iy$  must be inserted inside of disk.
 After performing Wick contractions, one finds
\beqa
{\cal A}^{C\phi}&=&
-(\frac{\pi\mu_p }{4})\xi_{1i}\Tr(P_{-}\fsH_{(n)}M_p\gamma^{i})
\label{CPrt}
\eeqa
The amplitude has been normalized
by multiplying $\frac{\sqrt{2} \,\pi }{4}\mu_p$. The normalization is chosen to match with the field theory computation. There will be similar rescalings for
other amplitudes below.
Above we have used the Jacobian
$j=|x_{14}x_{45}x_{51}|, \;\;x_4\equiv z, \;x_5\equiv\bar z$.
In the field theory, this amplitude is reproduced
 by the following coupling,
\beqa
S^{(1)}=\lambda\mu_p\int d^{p+1}\sigma {1\over (p+1)!}
(\veps^v)^{a_0\cdots a_{p}}\,\Tr\left(\Phi^i\right)\,
\partial_{i} C^{(p+1)}_{a_0\cdots a_{p}}(\sigma)\ .
 \label{Cphi}
\eeqa
The scalar field in the coupling above comes from the Taylor expansion (see section 5 of \cite{Hatefi:2012wj}). {Note also that $\mu_p$ is the RR charge of branes and $(\veps^v)^{a_0\cdots a_{p}}$ is the volume form
parallel to the world volume of the brane.}  Eq.\reef{Cphi}
 can be rewritten as
\beqa
S^{(1)}=\lambda\mu_p\int d^{p+1}\sigma {1\over (p+1)!}
(\veps^v)^{a_0\cdots a_{p}}\,\Tr\left(\Phi^i\right)\,
H^{(p+2)}_{ia_0\cdots a_{p}}(\sigma)\ .
\labell{cp431z2}
\eeqa
One can easily check that \reef{cp431z2} precisely reproduces the
string theory amplitude of \reef{CPrt}.
Our second example of a two-point amplitude is $<V_C V_A>$,
\begin{eqnarray}
{\cal A}^{CA} & \sim & \int dx_{1}dzd\bar{z}\,
  \lan V_{A}^{(-1)}{(x_{1})}
V_{RR}^{(-\frac{1}{2},-\frac{1}{2})}(z,\bar{z})\ran,\labell{cpsstring003}
\eeqa
 The final form of the amplitude is given by
\beqa {\cal A}^{CA}&\sim&
2^{-1/2}\xi_{1a}\Tr(P_{-}\fsH_{(n)}M_p\gamma^{a})(\frac{\pi\mu_p 2^{1/2}}{4})
\label{CP298r}
\eeqa
where the amplitude has been normalized with $\frac{\pi\mu_p 2^{1/2}}{4}$.
The result can be reproduced by field theory using a coupling between a RR ($p$-1)-form
and one gauge field; it takes the form of a Wess-Zumino coupling,
\beqa
S^{(2)}=(2\pi\alpha')\mu_p\int d^{p+1}\sigma
 C^{(p-1)}\wedge F\ .
\nonumber
\eeqa
 This WZ action can be rewritten as
\beqa
S^{(2)}=i(2\pi\alpha')\mu_p\int d^{p+1}\sigma {1\over (p)!}
(\veps^v)^{a_0\cdots a_{p}}
H^{(p)}_{a_0\cdots a_{p-1}}\xi_{a_{p}}\ .
\labell{cp241}
\eeqa
{Extracting the trace in \reef{CP298r} and  considering the coupling \reef{cp241} in field
theory exactly reproduce the amplitude of $<V_C V_A>$  in \reef{CP298r}.}

\subsection{ The three-point function $<V_C V_\phi V_\phi>$}

The scattering amplitude between one RR and two scalar fields $<V_CV_\phi V_\phi>$\footnote{
The three-point amplitude $<V_CV_A V_A>$ and its infinite contact terms can be found in \cite{Hatefi:2011jq} in the current context.}  is given by
\begin{eqnarray}
{\cal A}^{C\phi\phi} & \sim & \int dx_{1}dx_{2}dzd\bar{z}\,
  \lan V_{\phi}^{(-1)}{(x_{1})}
V_{\phi}^{(0)}{(x_{2})}
V_{RR}^{(-\frac{1}{2},
-\frac{1}{2})}(z,\bar{z})\ran,\labell{cppsstring}
\eeqa
After some algebra, one gets
\beqa
{\cal A}^{C\phi\phi}&\sim& \int
 dx_{1}dx_{2}dx_{4} dx_{5}\,
(P_{-}\fsH_{(n)}M_p)^{\al\be}\xi_{1i}\xi_{2j}x_{45}^{-1/4}(x_{14}x_{15})^{-1/2}\nonumber\\&&
\times(I_1+I_2)\,\labell{cpp125}
\eeqa
where $I_1$ and $I_2$ are given by
\beqa
I_1&=&{<:e^{\alpha'ik_1.X(x_1)}:\partial X^j(x_2)e^{\alpha'ik_2.X(x_2)}
:e^{i\frac{\alpha'}{2}p.X(x_4)}:e^{i\frac{\alpha'}{2}p.D.X(x_5)}:>}
 \  \non \\&&\times{<:S_{\al}(x_4):S_{\be}(x_5):\psi^i(x_1):>},\nonumber\\
I_2&=&{<:e^{\alpha'ik_1.X(x_1)}:e^{\alpha'ik_2.X(x_2)}
:e^{i\frac{\alpha'}{2}p.X(x_4)}:e^{i\frac{\alpha'}{2}p.D.X(x_5)}:>}
 \  \non \\&&\times{<:S_{\al}(x_4):S_{\be}(x_5)::\psi^i(x_1):\alpha'ik_{2a}\psi^{a}\psi^{j}(x_2)>}.
\label{cpp1234}
\eeqa
Using the basic OPEs
\cite{Liu:2001qa,Garousi:2007fk,Garousi:2008ge},
 the fermionic correlators are given by
\beqa
I_1^i&\equiv & <:S_{\al}(x_4):S_{\be}(x_5):\psi^i(x_1):>=2^{-1/2}x_{45}^{-3/4}(x_{14}x_{15})^{-1/2}
(\gamma^{i}C^{-1})_{\alpha\beta}.\label{1o2}
\eeqa
\beqa
<:S_{\al}(x_4):S_{\be}(x_5):\psi^a\psi^i(x_1):>&=& -\frac{1}{2}x_{45}^{-1/4}
x_{14}^{-1}x_{15}^{-1}(\Gamma^{ai}C^{-1})_{\al\be}.
\nonumber
\eeqa
Generalizing this, the correlation function of two spin operators, one fermion field
and  one current was
obtained in \cite{Hatefi:2010ik} according to which,
\beqa
I_2^{jai}&\equiv & <:S_{\al}(x_4):S_{\be}(x_5):\psi^i(x_1):\psi^a\psi^j(x_2):>\nonumber\\
&=&\bigg\{(\Gamma^{jai}C^{-1})_{\alpha\beta} +\frac{\alpha'
Re[x_{14}x_{25}]}{x_{12}x_{45}}\bigg(\eta^{ij}(\gamma^{a}C^{-1})_{\alpha\beta}\bigg)\bigg\}
\nonumber\\&&\times2^{-3/2}x_{45}^{1/4}(x_{24}x_{25})^{-1}(x_{14}x_{15})^{-1/2}.
\label{6cpp8}
 \eeqa
Substituting the spin correlators above into the amplitude and
working out the $X$ correlators, one finds
\beqa {\cal A}^{C\phi\phi}&\!\!\!\!\sim\!\!\!\!\!&\int dx_{1}dx_{2}
dx_{4}dx_{5}(P_{-}\fsH_{(n)}M_p)^{\al\be}I\xi_{1i}\xi_{2j}x_{45}^{-1/4}
(x_{14}x_{15})^{-1/2}\nonumber\\&&\times\bigg(I_1^i(a^j_1)+i\alpha'
k_{2a}I_2^{jai}\bigg)\labell{amp3cppp},\eeqa where \beqa
I&=&|x_{12}|^{ \alpha'^2k_1.k_2}|x_{14}x_{15}|^{\frac{\alpha'^2}{2}
k_1.p}|x_{24}x_{25}|^{\frac{\alpha'^2}{2} k_2.p}
|x_{45}|^{\frac{\alpha'^2}{4}p.D.p},\nonumber\\
a^j_1&=&ip^{j}\frac{x_{54}}{x_{24}x_{25}}. \label{icppsas} \eeqa
Let us gauge-fix $SL(2,R)$ invariance as follows:
\beqa (x_1,x_2,x_4,x_5)&=&(x,-x,i,-i)
\eeqa
with which the jacobian takes
$J=-2i(1+x^2)$. The amplitude now takes
\beqa {\cal A}^{C\phi\phi}&=&\int_{-\infty}^{\infty} dx (x^2+1)^{2t-1} x^{-2t} (2\xi_{1i} \xi_{2j} 2^{-1/2}2^{-2t})\nonumber\\&&\times
\bigg[-p^j\Tr(P_{-}\fsH_{(n)}M_p\gamma^{i})
+k_{2a}\Tr(P_{-}\fsH_{(n)}M_p\Gamma^{ija})\bigg]
\labell{mm55}
\eeqa
 where
$
t=\frac{-\alpha'}{2}(k_1+k_2)^2$. Note that the term $\frac{\alpha' Re[x_{14}x_{25}]}{x_{12}x_{45}}$ does
not contribute to the amplitude. This is because the
integrand is odd and the integration is over the whole worldsheet.
Carrying out the integration and using
momentum conservation, the final result of the
amplitude is shown to be
\beqa {\cal A}^{C\phi\phi}&=& (2\xi_{1i} \xi_{2j} 2^{-1/2}\pi^{1/2})\frac{\Gamma(-t+\frac{1}{2})}{\Gamma(-t+1)}\nonumber\\&&\times
\bigg[-p^j\Tr(P_{-}\fsH_{(n)}M_p\gamma^{i})
+k_{2a}\Tr(P_{-}\fsH_{(n)}M_p\Gamma^{ija})\bigg]
\labell{mm57}
\eeqa

\begin{center}
\begin{picture}
(600,120)(0,0)
\Line(195,35)(245,70)\Text(210,64)[]{$\phi_2$}
\Gluon(245,70)(245,115){4.20}{5.9}\Text(275,103)[]{$C_{p+1}$}
\Line(245,70)(295,35)\Text(285,65)[]{$\phi_1$}
\end{picture}\\ {\sl Figure 1 : The Feynman diagram corresponding to the amplitude of \reef{mm57}.}
\end{center}

Let us consider the low energy expansion in which the Mandelstam
variable $t$ is sent to $t\rightarrow 0$.
The limit is equivalent to taking $\al'\rightarrow 0$ on the string amplitudes.
 It turns out that the amplitude is non zero only for $n=p+2$.
It is obvious from the gamma functions that the desired amplitude has an infinite number of
contact terms.
The expansion of the $\Gamma$-function factors can be written as
\beqa
\sqrt{\pi}\frac{\Gamma(-t+\frac{1}{2})}{\Gamma(-t+1)}&=&\pi \sum_{n=-1}^{\infty} c_n(t)^{n+1}
\eeqa
The first few coefficients, $c_n$, are
\beqa
c_{-1}=1,\quad c_{0}=2ln(2),\quad c_{1}=\frac{\pi^2}{6}+2ln(2)^2
\eeqa
Let us find the SYM vertices that reproduce this string result.
First we start with a Chern-Simons action. The minimal form of
the vertex includes a bulk RR ($p$+1)-form potential and the two
world volume scalars, \beqa S^{(3)}
&=&i\frac{\lambda^2\mu_p}{2(p+1)!}\int d^{p+1}\sigma
(\veps^v)^{a_0\cdots a_{p}} \Tr(\Phi^i\Phi^j)
\prt_{i}\prt_{j}C^{(p+1)}_{a_0\cdots a_{p}}(\sigma) \nonumber\eeqa
The scalars have come from the Taylor expansion.\footnote{
Given a string amplitude that involves scalar vertices, there are three ways to construct
the corresponding field theory vertices. The first is through Wess-Zumino type interactions, and was proposed in the Myers'
paper \cite{Myers:1999ps}. The second is to examine pull-back procedure.
  The third - which we call "Taylor expansion" - was mentioned for example
  in section 5 of \cite{Hatefi:2012wj}. The terms of the third type take the form of Taylor expansion but
they do not arise from Wess-Zumino terms.
} One can rewrite the
coupling above as
\beqa S^{(3)} &=&i\frac{\lambda^2\mu_p}{2(p+1)!}\int d^{p+1}\sigma
(\veps^v)^{a_0\cdots a_{p}} \Tr(\Phi^i\Phi^j)
\prt_{j}H^{(p+2)}_{ia_0\cdots a_{p}}(\sigma) \label{S21} \eeqa
Normalizing the amplitude \reef{mm57} by $\frac{\pi \mu_p 2^{1/2}}{4}$ ,
it is possible to produce all the contact terms for the first term of
the amplitude \reef{mm57} by a higher derivative extension of the coupling above,
\beqa S^{(3)} &=&\frac{\lambda^2\mu_p}{2(p+1)!}\int d^{p+1}\sigma
(\veps^v)^{a_0\cdots a_{p}}p^j H^{(p+2)}_{ia_0\cdots
a_{p}}(\sigma)\nonumber\\&&\times
\sum_{n=-1}^{\infty}c_{n}(\alpha')^{n+1}\Tr(\prt_{a_{1}}...\prt_{a_{n+1}}\Phi^i\prt^{a_{1}}...\prt^{a_{n+1}}\Phi^j)
\label{S22} \eeqa
where $H^{(p+2)}=dC^{(p+1)}$. Now we are going to re-derive all infinite contact
terms of the second term of \reef{mm57}. In order to do so,
  the following interaction vertex must be taken as well,
\beqa S^{(4)} &=&i \frac{\lambda^2\mu_p} {2(p)!}\int d^{p+1}\sigma
(\veps^v)^{a_0\cdots a_{p}} \Tr(D_{a_{0}}\Phi^iD_{a_{1}}\Phi^j)(p)
C^{(p+1)}_{ija_2\cdots a_{p}}(\sigma) \label{pull} \eeqa
The scalars come from pull-back.
By considering the antisymmetric property of $(\veps^v)^{a_0\cdots
a_{p}}$, the higher derivative extension of \reef{pull} can be
re-expressed as \beqa S^{(4)} &=&i \frac{\lambda^2\mu_p} {2(p)!}\int
d^{p+1}\sigma (\veps^v)^{a_0\cdots a_{p}}
\sum_{n=-1}^{\infty}c_{n}(\alpha')^{n+1}\Tr(\prt_{a_{1}}...\prt_{a_{n+1}}
D_{a_{0}}\Phi^i \prt^{a_{1}}...\prt^{a_{n+1}} \Phi^j)
 H^{(p+2)}_{ija_1\cdots a_{p}}(\sigma)
\nonumber
\eeqa
Since only the couplings between one RR and two scalar fields are
relevant, one can replace the covariant derivatives on the scalar
fields by their partial derivatives.

\section{Universality and analysis of $<V_CV_\phi V_\phi V_\phi>$ \label{main}}

In this section, we compute $<V_CV_\phi V_\phi V_\phi>$\footnote{
Due to the presence of a closed string state, the computation shares
the technical aspects of a pure open string five-point amplitude (see, e.g., \cite{Medina:2002nk}). } and analyze its infinite poles
and some of its contact terms.
The corresponding low energy field theory vertices are determined by universality,
and are subsequently shown to match the field theory poles and contact terms with those of the
string amplitude. Because of the similarities between the present
amplitudes and the amplitudes that were considered in the previous
works, the present analysis shares some parts of the
computations with the earlier works. However, most of
final results concerning contact interactions cannot be derived, for example,
by applying T-duality to the previous results as we will discuss in
the main body.

\vspace{.2in}

More specifically, the amplitude of
$<V_C V_A V_A V_A>$ was analyzed in \cite{Hatefi:2010ik}, and one might wonder whether the
amplitude of $<V_C V_\f V_\f V_\f>$ could be derived by applying
T-duality. After computing $<V_C V_\f V_\f V_\f>$, we will take up
this issue. For example, we will see that the amplitude of $<V_C V_\f V_\f V_\f>$
has some extra terms that are absent in $<V_C V_A V_A V_A>$.

\vspace{.1in}

 As a comment we were not able to reproduce all
contact terms of the four-point amplitude $<V_C V_\f V_\f V_\f>$  with usual pull-back. This is a hint that pull-back must be modified
\cite{Hull:1997jc},\cite{Dorn:1996xk}.

\vspace{.1in}
Although we could find the field theory vertices that
reproduce the entire pole terms and some of the contact terms of the
string result for our four-point function $<V_CV_\phi V_\phi V_\phi>$, more work
is required\footnote{
 It might be a
hint that pull-back may need to be modified.
We will discuss this issue in the conclusion section.
} to find the vertices that would reproduce all
contact terms.

\subsection{Computation of $<V_CV_\phi V_\phi V_\phi>$}

Let us turn to our main result, the analysis of the amplitude
of one closed string RR field and three scalar fields,
\begin{eqnarray}
{\cal A}^{C\phi\phi\phi} & \sim & \int dx_{1}dx_{2}dx_{3}dzd\bar{z}\,
  \lan V_{\phi}^{(-1)}{(x_{1})}
V_{\phi}^{(0)}{(x_{2})}V_\phi^{(0)}{(x_{3})}
V_{RR}^{(-\frac{1}{2},-\frac{1}{2})}(z,\bar{z})
\ran,\labell{sstring43}
\eeqa
To obtain a precise result we must apply
correct Wick and modified
Wick-like contraction (see \cite {Hatefi:2010ik,Hatefi:2012wj}).
With various Wick contractions, this amplitude reduces to
\beqa {\cal A}^{C\phi\phi\phi}&\sim& \int
 dx_{1}dx_{2}dx_{3}dx_{4} dx_{5}\,
(P_{-}\fsH_{(n)}M_p)^{\al\be}\xi_{1i}\xi_{2j}\xi_{3k}x_{45}^{-1/4}(x_{14}x_{15})^{-1/2}\nonumber\\&&
\times(I_1+I_2+I_3+I_4)\Tr(\lam_1\lam_2\lam_3),\labell{125}
\eeqa
for a particular ordering that we call 123 ordering.
 The explicit expressions for $I$'s are
\beqa I_1&=&{<:e^{\alpha'ik_1.X(x_1)}:\partial
X^j(x_2)e^{\alpha'ik_2.X(x_2)} :\partial
X^k(x_3)e^{\alpha'ik_3.X(x_3)}:e^{\frac{\alpha'}{2}ip.X(x_4)}:e^{\frac{\alpha'}{2}ip.D.X(x_5)}:>}
 \  \non \\&&\times{<:S_{\al}(x_4):S_{\be}(x_5):\psi^i(x_1):>},\nonumber\\
I_2&=&{<:e^{\alpha'ik_1.X(x_1)}:e^{\alpha'ik_2.X(x_2)} :\partial
X^k(x_3)e^{\alpha'ik_3.X(x_3)}:e^{\frac{\alpha'}{2}ip.X(x_4)}:e^{\frac{\alpha'}{2}ip.D.X(x_5)}:>}
 \  \non \\&&\times{<:S_{\al}(x_4):S_{\be}(x_5)::\psi^i(x_1):\alpha'ik_2.\psi\psi^{j}(x_2)>},\nonumber\\
 I_3&=&{<: e^{\alpha'ik_1.X(x_1)}:\partial X^j(x_2)e^{\alpha'ik_2.X(x_2)}
:e^{\alpha'ik_3.X(x_3)}:e^{\frac{\alpha'}{2}ip.X(x_4)}:e^{\frac{\alpha'}{2}ip.D.X(x_5)}:>}
 \  \non \\&&\times{<:S_{\al}(x_4):S_{\be}(x_5)::\psi^i(x_1):\alpha'ik_3.\psi\psi^{k}(x_3)>},\nonumber\\
 I_4&=&{<: e^{\alpha'ik_1.X(x_1)}:e^{\alpha'ik_2.X(x_2)}
:e^{\alpha'ik_3.X(x_3)}:e^{\frac{\alpha'}{2}ip.X(x_4)}:e^{\frac{\alpha'}{2}ip.D.X(x_5)}:>}
 \  \non \\&&\times{<:S_{\al}(x_4):S_{\be}(x_5):\psi^i(x_1)
:\alpha'ik_{2}\cd\psi\psi^j(x_2):\alpha'ik_{3}\cd\psi\psi^k(x_3):>}.
\label{i1234}
\eeqa
 Using the results of \cite{Garousi:2008ge,Hatefi:2012wj,Hatefi:2012ve},
one can show that
\beqa
I_5^{jai}&=&<:S_{\al}(x_4):S_{\be}(x_5):\psi^i(x_1):\psi^a\psi^j(x_2):>\nonumber\\
&=&\bigg\{(\Gamma^{jai}C^{-1})_{\alpha\beta}
+\frac{\alpha' Re[x_{14}x_{25}]}{x_{12}x_{45}}\bigg(\eta^{ij}(\gamma^{a}C^{-1})_{\alpha\beta}\bigg)\bigg\}
\nonumber\\&&\times2^{-3/2}x_{45}^{1/4}(x_{24}x_{25})^{-1}(x_{14}x_{15})^{-1/2}.
\label{68}
\eeqa
The correct result for the correlation function
between two spin operators, two  currents and one worldsheet
fermion was obtained in \cite{Hatefi:2010ik},
\beqa
I_6^{kbjai}&=&<:S_{\al}(x_4):S_{\be}(x_5)::\psi^i(x_1):\psi^a\psi^j(x_2):\psi^b\psi^k(x_3)>\nonumber\\
&=&\bigg\{(\Gamma^{kbjai}C^{-1})_{{\alpha\beta}}+\alpha' r_1\frac{Re[x_{14}x_{25}]}{x_{12}x_{45}}+\alpha' r_2\frac{Re[x_{14}x_{35}]}{x_{13}x_{45}}+\alpha' r_3\frac{Re[x_{24}x_{35}]}{x_{23}x_{45}}+\alpha'^2 r_4\nonumber\\&&\times\bigg(\frac{Re[x_{24}x_{35}]}{x_{23}x_{45}}\bigg)^{2}
+\alpha'^2 r_5\bigg(\frac{Re[x_{14}x_{25}]}{x_{12}x_{45}}\times\frac{Re[x_{24}x_{35}]}{x_{23}x_{45}}\bigg)+\alpha'^2 r_6\bigg(\frac{Re[x_{14}x_{35}]}{x_{13}x_{45}}\nonumber\\&&\times\frac{Re[x_{24}x_{35}]}{x_{23}x_{45}}\bigg)
\bigg\}2^{-5/2}x_{45}^{5/4}(x_{24}x_{25}x_{34}x_{35})^{-1}(x_{14}x_{15})^{-1/2},\label{hh33}
\eeqa
where
\beqa
r_1&=&\bigg(\eta^{ij}(\Gamma^{kba}C^{-1})_{\alpha\beta}\bigg),\nonumber\\
r_2&=&\bigg(\eta^{ik}(\Gamma^{bja}C^{-1})_{\alpha\beta}
\bigg),\nonumber\\
r_3&=&\bigg(\eta^{ab}(\Gamma^{kji}C^{-1})_{\alpha\beta}
+\eta^{jk}(\Gamma^{bai}C^{-1})_{\alpha\beta}\bigg),\nonumber\\
r_4&=&\bigg((-\eta^{ab}\eta^{jk})(\gamma^{i}C^{-1})_{\alpha\beta}\bigg),\nonumber\\
r_5&=&\bigg((-\eta^{ji}\eta^{ab})(\gamma^{k}C^{-1})_{\alpha\beta}\bigg),\nonumber\\
r_6&=&\bigg((\eta^{ik}\eta^{ab})(\gamma^{j}C^{-1})_{\alpha\beta} \bigg).
\eeqa
Inserting the spin correlators above in the amplitude and performing contractions over $X$, one finds:
\beqa
{\cal A}^{C\phi\phi\phi}&\!\!\!\!\sim\!\!\!\!\!&\int dx_{1}dx_{2} dx_{3}dx_{4}dx_{5}(P_{-}\fsH_{(n)}M_p)^{\al\be}I\xi_{1i}\xi_{2j}\xi_{3k}x_{45}^{-1/4}(x_{14}x_{15})^{-1/2}\nonumber\\&&\times\bigg(I_7^i(-\eta^{jk}x_{23}^{-2}+a^j_1a^k_2)+a^k_2a^{ji}_3+a^j_1a^{ki}_4-\alpha'^2 k_{2a}k_{3b}I_6^{kbjai}\bigg)\Tr(\lam_1\lam_2\lam_3)
\labell{amp3q},
\eeqa
where  $I_6^{kbjai}$ is given in \reef{hh33}, and
\beqa I&=&|x_{12}|^{ \alpha'^2k_1.k_2}|x_{13}|^{\alpha'^2
k_1.k_3}|x_{14}x_{15}|^{\frac{\alpha'^2}{2}
k_1.p}|x_{23}|^{\alpha'^2
k_2.k_3}|x_{24}x_{25}|^{\frac{\alpha'^2}{2} k_2.p}
|x_{34}x_{35}|^{\frac{\alpha'^2}{2} k_3.p}|x_{45}|^{\frac{\alpha'^2}{4}p.D.p},\nonumber\\
a^j_1&=&ip^{j}\frac{x_{54}}{x_{24}x_{25}},\nonumber\\
a^k_2&=&ip^{k}\frac{x_{54}}{x_{34}x_{35}},\nonumber\\
a^{ji}_3&=&\alpha' ik_{2a}I_5^{jai},\nonumber\\
a^{ki}_4&=&\alpha' ik_{3b}2^{-3/2}x_{45}^{1/4}(x_{34}x_{35})^{-1}(x_{14}x_{15})^{-1/2} \nonumber\\&&\times\bigg\{(\Gamma^{kbi}C^{-1})_{\alpha\beta}
+\frac{\alpha' Re[x_{14}x_{35}]}{x_{13}x_{45}}\bigg(\eta^{ik}(\gamma^{b}C^{-1})_{\alpha\beta}\bigg)\bigg\}
,\nonumber\\
I_7^i&=&<:S_{\al}(x_4):S_{\be}(x_5):\psi^i(x_1):>=2^{-1/2}x_{45}^{-3/4}(x_{14}x_{15})^{-1/2}
(\gamma^{i}C^{-1})_{\alpha\beta}.
\label{isas}
\eeqa
Using the integral presented in, e.g.,
\cite{Fotopoulos:2001pt} and \cite{Hatefi:2012wj},
 one can rewrite the amplitude \reef{amp3q} as
\beqa {\cal A}^{C\phi\phi\phi }&=&{\cal A}_{1}+{\cal A}_{2}+{\cal A}_{3}+{\cal A}_{4}+{\cal A}_{5}+{\cal A}_{6}
+{\cal A}_{7}+{\cal A}_{8}+{\cal A}_{9}+{\cal A}_{10}\labell{11u}
\eeqa
with
\beqa
{\cal A}_{1}&\!\!\!\sim\!\!\!&-2^{-1/2}\xi_{1i}\xi_{2j}\xi_{3k}
\bigg[k_{3b}k_{2a}\Tr(P_{-}\fsH_{(n)}M_p\Gamma^{kbjai})-k_{3b}p^j\Tr(P_{-}\fsH_{(n)}M_p\Gamma^{kbi})\nonumber\\&&-k_{2a}p^k\Tr(P_{-}\fsH_{(n)}M_p\Gamma^{jai})+p^jp^k\Tr(P_{-}\fsH_{(n)}M_p\gamma^{i})\bigg]
L_1,
\nonumber\\
{\cal A}_{2}&\sim&2^{-1/2}
\bigg\{
2\xi_{1}.\xi_{2}k_{2a}k_{3b}\xi_{3k}\Tr(P_{-}\fsH_{(n)}M_p \Gamma^{kba})\bigg\}L_2\nonumber\\
{\cal A}_{3}&\sim&2^{-1/2}
\bigg\{\xi_{1i}\xi_{2j}\xi_{3k}\Tr(P_{-}\fsH_{(n)}M_p \Gamma^{kji})\bigg\}L_{22}\nonumber\\
{\cal A}_{4}&\sim&-2^{-1/2}
\bigg\{
2\xi_{3}.\xi_{1}k_{2a}k_{3b}\xi_{2j}\Tr(P_{-}\fsH_{(n)}M_p \Gamma^{bja})\bigg\}L_4\nonumber\\
\nonumber\\
{\cal A}_{5}&\sim&2^{-1/2}
\bigg\{2\xi_{3}.\xi_{2}k_{2a}k_{3b}\xi_{1i}\Tr(P_{-}\fsH_{(n)}M_p \Gamma^{bai})\bigg\}L_5\nonumber\\
{\cal A}_{6}&\sim&2^{1/2}L_{33}\bigg\{-k_{2a}p^k\xi_1.\xi_2\xi_{3k}\Tr(P_{-}\fsH_{(n)}M_p\gamma^a)
\bigg\}
\nonumber\\
{\cal A}_{7}&\sim&2^{1/2}L_3\bigg\{
k_{3b}p^j\xi_1.\xi_3\xi_{2j}\Tr(P_{-}\fsH_{(n)}M_p\gamma^b)\bigg\}
\nonumber\\
{\cal A}_{8}&\sim&2^{1/2}L_6\bigg\{\xi_{2j}\Tr(P_{-}\fsH_{(n)}M_p\gamma^j)
(ut\xi_1.\xi_3)\bigg\}.
\nonumber\\
{\cal A}_{9}&\sim&2^{1/2}L_6\bigg\{\xi_{3k}\Tr(P_{-}\fsH_{(n)}M_p\gamma^k)
(us\xi_1.\xi_2)\bigg\}
\nonumber\\
{\cal A}_{10}&\sim&2^{1/2}L_6\bigg\{\xi_{1i}\Tr(P_{-}\fsH_{(n)}M_p\gamma^i)
(ts\xi_3.\xi_2)\bigg\}
\labell{480}
\eeqa
where
\beqa
L_1&=&(2)^{-2(t+s+u)+1}\pi{\frac{\Gamma(-u+\frac{1}{2})
\Gamma(-s+\frac{1}{2})\Gamma(-t+\frac{1}{2})\Gamma(-t-s-u+1)}
{\Gamma(-u-t+1)\Gamma(-t-s+1)\Gamma(-s-u+1)}},\nonumber\\
L_2&=&(2)^{-2(t+s+u)}\pi{\frac{\Gamma(-u+1)
\Gamma(-s+1)\Gamma(-t)\Gamma(-t-s-u+\frac{1}{2})}
{\Gamma(-u-t+1)\Gamma(-t-s+1)\Gamma(-s-u+1)}}
\nonumber\\
L_{22}&=&(2)^{-2(t+s+u)}\pi{\frac{\Gamma(-u+1)
\Gamma(-s+1)\Gamma(-t+1)\Gamma(-t-s-u+\frac{1}{2})}
{\Gamma(-u-t+1)\Gamma(-t-s+1)\Gamma(-s-u+1)}}
\nonumber\\
L_{33}&=&(2)^{-2(t+s+u)}\pi{\frac{\Gamma(-u+1)
\Gamma(-s+1)\Gamma(-t)\Gamma(-t-s-u+\frac{1}{2})}{\Gamma(-u-t+1)
\Gamma(-t-s+1)\Gamma(-s-u+1)}}
\nonumber\\
L_3&=&(2)^{-2(t+s+u)}\pi{\frac{\Gamma(-u+1)
\Gamma(-s)\Gamma(-t+1)\Gamma(-t-s-u+\frac{1}{2})}{\Gamma(-u-t+1)
\Gamma(-t-s+1)\Gamma(-s-u+1)}}
,\nonumber\\
L_4&=&(2)^{-2(t+s+u)}\pi{\frac{\Gamma(-u+1)
\Gamma(-s)\Gamma(-t+1)\Gamma(-t-s-u+\frac{1}{2})}
{\Gamma(-u-t+1)\Gamma(-t-s+1)\Gamma(-s-u+1)}}
,\nonumber\\
L_5&=&(2)^{-2(t+s+u)}\pi{\frac{\Gamma(-u)
\Gamma(-s+1)\Gamma(-t+1)\Gamma(-t-s-u+\frac{1}{2})}
{\Gamma(-u-t+1)\Gamma(-t-s+1)\Gamma(-s-u+1)}}
,\nonumber\\
L_6&=&(2)^{-2(t+s+u)-1}\pi{\frac{\Gamma(-u+\frac{1}{2})
\Gamma(-s+\frac{1}{2})\Gamma(-t+\frac{1}{2})\Gamma(-t-s-u)}
{\Gamma(-u-t+1)\Gamma(-t-s+1)\Gamma(-s-u+1)}},
\label{Ls}
\eeqa
where
\beqa
s&=&\frac{-\alpha'}{2}(k_1+k_3)^2,\quad t=\frac{-\alpha'}{2}(k_1+k_2)^2,\quad u=\frac{-\alpha'}{2}(k_2+k_3)^2
\eeqa
${\cal A}^{C\phi\phi\phi }$ can be further simplified to
\beqa {\cal A}^{C\phi\phi\phi}&=&{\cal A}'_{1}+{\cal A}'_{2}
 +{\cal A}'_{3},\labell{141u}
\eeqa
where
\beqa
{\cal A}'_{1}&\!\!\!\sim\!\!\!&2^{-1/2}2\xi_{1i}\xi_{2j}\xi_{3k}(t+s+u)L_1'
\bigg[k_{3b}k_{2a}\Tr(P_{-}\fsH_{(n)}M_p\Gamma^{kbjai})-k_{3b}p^j\Tr(P_{-}\fsH_{(n)}M_p\Gamma^{kbi})\nonumber\\&&-k_{2a}p^k\Tr(P_{-}\fsH_{(n)}M_p\Gamma^{jai})+p^jp^k\Tr(P_{-}\fsH_{(n)}M_p\gamma^{i})\bigg]
\nonumber\\
{\cal A}'_{2}&\sim&2^{-1/2}L_2'
\bigg\{2us\xi_{1}.\xi_{2}k_{2a}k_{3b}\xi_{3k}\Tr(P_{-}\fsH_{(n)}M_p \Gamma^{kba})-ust\xi_{1i}\xi_{2j}\xi_{3k}\Tr(P_{-}\fsH_{(n)}M_p \Gamma^{kji})
\nonumber\\&&
-2ut\xi_{3}.\xi_{1}k_{2a}k_{3b}\xi_{2j}\Tr(P_{-}\fsH_{(n)}M_p \Gamma^{bja})+2st\xi_{3}.\xi_{2}k_{2a}k_{3b}\xi_{1i}\Tr(P_{-}\fsH_{(n)}M_p \Gamma^{bai})
\nonumber\\&&
-2usk_{2a}p^k\xi_1.\xi_2\xi_{3k}\Tr(P_{-}\fsH_{(n)}M_p\gamma^a)
+2utk_{3b}p^j\xi_1.\xi_3\xi_{2j}\Tr(P_{-}\fsH_{(n)}M_p\gamma^b)\bigg\}\nonumber\\
{\cal A}'_{3}&\sim&2^{-1/2}L_1'\bigg\{\bigg[\xi_{1i}\Tr(P_{-}\fsH_{(n)}M_p\gamma^i)
(ts\xi_3.\xi_2)\bigg]+\bigg[1\leftrightarrow 2\bigg]
+\bigg[1\leftrightarrow 3\bigg]\bigg\}.
\labell{4798}
\eeqa
The functions $L_1',L_2'$ are given as follows:
\beqa
L_1'&=&(2)^{-2(t+s+u)}\pi{\frac{\Gamma(-u+\frac{1}{2})
\Gamma(-s+\frac{1}{2})\Gamma(-t+\frac{1}{2})\Gamma(-t-s-u)}
{\Gamma(-u-t+1)\Gamma(-t-s+1)\Gamma(-s-u+1)}},\nonumber\\
L_2'&=&(2)^{-2(t+s+u)}\pi{\frac{\Gamma(-u)
\Gamma(-s)\Gamma(-t)\Gamma(-t-s-u+\frac{1}{2})}{\Gamma(-u-t+1)
\Gamma(-t-s+1)\Gamma(-s-u+1)}}
\label{L1L2}
\eeqa
{For the amplitude of two fermions and three massless scalar
vertex operators, there is no correlation between $<\partial {X^j}(x_2) e^{2ik.X(x)}>$ and $<\partial {X^k}(x_3) e^{2ik.X(x)}>$; one can easily see that
the terms ${\cal A}_{6},{\cal
A}_{7}$, and all terms of ${\cal A}_{1}$ except its first term identically vanish.}
Since $\fsH_{(n)}, M_p$ and $\Gamma^{kji}$
 are totally antisymmetric combinations of the
Gamma matrices, it follows that the amplitude is nonzero only for
$n=p+4,\; n=p+2$, and $p=n$. From the poles of the gamma functions,
one can easily see that the scattering amplitude has infinite
massless  and massive poles. In order to
compare this with the field theory, which has massless fields, one
must expand the amplitude so that the  massless poles of
the field theory remain while all massive poles rearrange themselves
in the form of contact interactions. The low energy expansion is
carried out by sending all Mandelstam variables to zero.

 As stated in the introduction, the above results cannot be entirely derived by applying T-duality
 to the previous result that was obtained for $<V_C V_A V_A V_A>$ in \cite{Hatefi:2010ik}.
 All the terms that contain the transverse components of the momentum $p^i$ such as $p^jp^k$, the second and third terms in ${\cal A}_1'$, and also the last two terms in ${\cal A}_2'$ in \reef{4798}, are not present in
 the corresponding result of \cite{Hatefi:2010ik}. We will ponder this issue in the conclusion.

\vspace{.2in}

\noindent Let us momentum-expand the string amplitude above in the $\alpha'\rightarrow 0$ limit,
\beqa
s\rightarrow 0,\qquad t\rightarrow 0,\qquad  u\rightarrow 0. \labell{point}
\eeqa
The Mandelstam variables satisfy\footnote{ The constraint
\reef{cons} implies that $p_ap^a\rightarrow 0$. Also note that it is
known that for  amplitudes including tachyon and RR, the expansion
makes sense only for a constant value of $p^ap_a\rightarrow
\frac{1}{4}$ \cite {Hatefi:2012wj}.
}
\beqa
s+t+u=-p_ap^a.
\labell{cons}
\eeqa
Note that the combination $ st L_1'$ has appeared in eq.(\ref{4798}) and that $L_1'$ is symmetric
 under the interchange of $(u,t,s)$. For the $st\, L_1'$ term, the proper expansion  is
\beqa
st L_1'&=&-{\pi^{5/2}}\left( \sum_{n=0}^{\infty}c_n(s+t+u)^n\right.
\left.+\frac{\sum_{n,m=0}^{\infty}c_{n,m}[s^n t^m +s^m t^n]}{(t+s+u)}\right.\nonumber\\
&&\left.+\sum_{p,n,m=0}^{\infty}f_{p,n,m}(s+t+u)^p[(s+t)^{n}(st)^{m}]\right),\label{expansion11}
\eeqa
When considering $su L_1'$, the proper expansion for $su L_1'$ is such that $t\leftrightarrow u$ in the expansion above. Similarly when
considering $tu L_1'$, the proper expansion for  $tu L_1'$ is such that $s\leftrightarrow u$ in \reef{expansion11}.
The expansion for $L_2'$ can be similarly summarized:
\beqa
su L_2'&=&-\pi^{3/2}\sum_{n=-1}^{\infty}b_n\bigg(\frac{1}{t}(u+s)^{n+1}\bigg)
            +\sum_{p,n,m=0}^{\infty}e_{p,n,m}t^{p}(su)^{n}(s+u)^m
\nonumber\\
st L_2'&=&-\pi^{3/2}\sum_{n=-1}^{\infty}b_n\bigg(\frac{1}{u}(t+s)^{n+1}\bigg)
        +\sum_{p,n,m=0}^{\infty}e_{p,n,m}u^{p}(ts)^{n}(s+t)^m
\nonumber\\
tu L_2'&=&-\pi^{3/2}\sum_{n=-1}^{\infty}b_n\bigg(\frac{1}{s}(u+t)^{n+1}\bigg)
             +\sum_{p,n,m=0}^{\infty}e_{p,n,m}s^{p}(tu)^{n}(t+u)^m
\labell{expansion44}
\eeqa
where some of the coefficients $b_n,\,e_{p,n,m},\,c_n,\,c_{n,m}$ and $f_{p,n,m}$ are
\beqa
&&b_{-1}=1,\,b_0=0,\,b_1=\frac{1}{6}\pi^2,\,b_2=2\z(3),c_0=0,c_1=-\frac{\pi^2}{6}\labell{hash},\\
&&e_{2,0,0}=e_{0,1,0}=2\z(3),e_{1,0,0}=\frac{1}{6}\pi^2,e_{1,0,2}=\frac{19}{60}\pi^4,e_{1,0,1}=e_{0,0,2}=6\z(3),\nonumber\\
&&c_2=-2\z(3),
\,c_{1,1}=\frac{\pi^2}{6},\,c_{0,0}=\frac{1}{2},c_{3,1}=c_{1,3}=\frac{2}{15}\pi^4,c_{2,2}=\frac{1}{5}\pi^4,\nonumber\\
&&c_{1,0}=c_{0,1}=0,
c_{3,0}=c_{0,3}=0\,
,\,c_{2,0}=c_{0,2}=\frac{\pi^2}{6},c_{1,2}=c_{2,1}=-4\z(3),c_{4,0}=c_{0,4}=\frac{1}{15}\pi^4 ,\nonumber
\eeqa
The coefficients $b_n$ appeared in the momentum expansion of the
S-matrix element of one RR, three gauge field vertex operators
\cite{Hatefi:2010ik}. The function of ${ L}_1'$ has infinite massless
scalar poles in the $(t+s+u)$-channel (in contrast with the amplitude of $<V_CV_A V_A V_A>$, which has infinite gauge but not scalar poles) and
${ L}_2'$ has infinite massless gauge poles in  $t$-,$s$-and $u$-channels.

\subsection{Universality in all-$\alpha'$ order higher derivative corrections of non-BPS and BPS branes}

Below we determine all higher derivative {corrections of non-BPS and BPS branes by utilizing} the
pattern that appeared in previous works that we now review.

Our first example of the pattern is the case of two tachyons and
two scalar fields on the world volume of $N$ non-BPS D-branes \cite{Hatefi:2012wj}. Given the
leading order vertices as follows
 \beqa 2T_p(\pi\alpha')^3{\rm
STr} \left(
\frac{}{}m^2\cT^2(D_a\phi^iD^a\phi_i)+\frac{}{}D^{\alpha}\cT
D_{\alpha}\cT D_a\phi^iD^a\phi_i- 2D_a\phi^iD_b\phi_i D^{b}\cT
D^{a}\cT \right)\labell{dbicoupling}
\eeqa
the all-order vertices turned out to be
 \beqa
\cL_{}&=&-2T_p(\pi\alpha')(\alpha')^{2+n+m}\sum_{n,m=0}^{\infty}(\cL_{1}^{nm}+\cL_{2}^{nm}+\cL_{3}^{nm}+\cL^{nm}_{4})\labell{lagrang}\eeqa
where \beqa \cL_1^{nm}&=&m^2
\Tr\left(\frac{}{}a_{n,m}[\cD_{nm}(\cT^2 D_a\phi^iD^a\phi_i)+ \cD_{nm}(D_a\phi^iD^a\phi_i\cT^2)]\right.\nonumber\\
&&\left.+\frac{}{}b_{n,m}[\cD'_{nm}(\cT D_a\phi^i\cT D^a\phi_i)+\cD'_{nm}( D_a\phi^i\cT D^a\phi_i\cT)]+h.c.\right)\nonumber\\
\cL_2^{nm}&=&\Tr\left(\frac{}{}a_{n,m}[\cD_{nm}(D^{\alpha}\cT D_{\alpha}\cT D_a\phi^iD^a\phi_i)+\cD_{nm}( D_a\phi^iD^a\phi_i D^{\alpha}\cT D_{\alpha}\cT)]\right.\nonumber\\
&&\left.+\frac{}{}b_{n,m}[\cD'_{nm}(D^{\alpha} \cT D_a\phi^i D_{\alpha}\cT D^a\phi_i)+\cD'_{nm}( D_a\phi^i D_{\alpha}\cT D^a\phi_i D^{\alpha} \cT)]+h.c.\right)\nonumber\\
\cL_3^{nm}&=&-\Tr
\left(\frac{}{}a_{n,m}[\cD_{nm}(D^{\beta}\cT D_{\mu}\cT D^\mu\phi^iD_\beta\phi_i)+\cD_{nm}( D^\mu\phi^iD_\beta\phi_iD^{\beta}\cT D_{\mu}\cT)]\right.\nonumber\\
&&\left.+\frac{}{}b_{n,m}[\cD'_{nm}(D^{\beta}\cT D^\mu\phi^iD_{\mu}\cT D_\beta\phi_i)+\cD'_{nm}(D^\mu\phi^i D_{\mu}\cT  D_\beta\phi_i  D^{\beta}\cT)]+h.c.\right)\nonumber\\
\cL_4^{nm}&=&-\Tr\left(\frac{}{}a_{n,m}[\cD_{nm}(D^{\beta}\cT
D^{\mu}\cT D_\beta\phi^iD_\mu\phi_i)
+\cD_{nm}( D^\beta\phi^iD^\mu\phi_iD_{\beta}\cT D_{\mu}\cT)]\right.\nonumber\\
&&\left.+\frac{}{}b_{n,m}[\cD'_{nm}(D^{\beta}\cT
D_\beta\phi^iD^{\mu}\cT D_\mu\phi_i)+\cD'_{nm}( D_\beta\phi^i
D_{\mu}\cT  D^\mu\phi_i D^{\beta}\cT)]+h.c. \right)\nonumber
\eeqa
where
\beqa
\cD_{nm}(EFGH)&\equiv&D_{b_1}\cdots D_{b_m}D_{a_1}\cdots D_{a_n}E  F D^{a_1}\cdots D^{a_n}GD^{b_1}\cdots D^{b_m}H,\nonumber\\
\cD'_{nm}(EFGH)&\equiv&D_{b_1}\cdots D_{b_m}D_{a_1}\cdots D_{a_n}E
D^{a_1}\cdots D^{a_n}F G D^{b_1}\cdots D^{b_m}H.\label{esiesi1}
\eeqa
The crucial step seems to extract the symmetric trace in terms of the ordinary trace and apply the
  higher derivative corrections $\cD_{nm},\cD'_{nm}$ on it.

 The second example is the amplitude of two tachyons and two gauge fields on the worldvolume of non-BPS branes \cite{Garousi:2008ge}.
One just applies $\cD_{nm},\cD'_{nm}$ on the couplings between two tachyons and two gauge fields.\footnote{The prescription works for brane anti-brane systems as well.
The only subtlety in this case is that after applying $\cD_{nm},\cD'_{nm}$, all $b_{n,m}$ must be  rendered to $-b_{n,m}$ \cite{Hatefi:2012eh111}.}
Now let us turn to BPS systems.

\vspace{.2in}

The third example is four-gauge field couplings
\beqa -T_p (2\pi\alpha')^4
S\Tr\left(-\frac{1}{8}F_{bd}F^{df}F_{fh}F^{hb}+\frac{1}{32}(F_{ab}F^{ba})^2\right).
\label{55} \eeqa
The closed form of the higher
derivative corrections of four-gauge fields to all orders of
$\alpha'$ (which must be added to DBI) was shown \cite{Hatefi:2010ik} to be
\beqa
(2\pi\alpha')^4\frac{1}{8\pi^2}T_p\left(\alpha'\right)^{n+m}\sum_{m,n=0}^{\infty}(\cL_{5}^{nm}+\cL_{6}^{nm}+\cL_{7}^{nm}),\labell{highder}\eeqa
with \beqa &&\cL_{5}^{nm}=-
\Tr\left(\frac{}{}a_{n,m}\cD_{nm}[F_{bd}F^{df}F_{fh}F^{hb}]+\frac{}{} b_{n,m}\cD'_{nm}[F_{bd}F_{fh}F^{df}F^{hb}]+h.c.\frac{}{}\right),\nonumber\\
&&\cL_{6}^{nm}=-\Tr\left(\frac{}{}a_{n,m}\cD_{nm}[F_{bd}F^{df}F^{hb}F_{fh}]+\frac{}{}b_{n,m}\cD'_{nm}[F_{bd}F^{hb}F^{df}F_{fh}]+h.c.\frac{}{}\right),\nonumber\\
&&\cL_{7}^{nm}=\frac{1}{2}\Tr\left(\frac{}{}a_{n,m}\cD_{nm}[F_{ab}F^{ab}F_{cd}F^{cd}]+\frac{}{}b_{n,m}\cD'_{nm}[F_{ab}F^{cd}F^{ab}F_{cd}]+h.c\frac{}{}\right),\nonumber\eeqa
where the higher derivative operators $D_{nm} $ and $ D'_{nm}$ are
defined in \reef{esiesi1}.
These couplings are exact up to total derivative terms and these
corrections have been checked by explicit computations of the
amplitude of one RR and three gauge fields \cite{Hatefi:2010ik}.

The fourth example \cite{Hatefi:2012wj} is two scalar and two
gauge field couplings,
\beqa &&- \frac{T_p(2\pi\alpha')^4}{2}{\rm STr}
\left(D_a\phi^iD^b\phi_iF^{ac}F_{bc}-\frac{1}{4} (D_a\phi^i
D^a\phi_iF^{bc}F_{bc})\right).\labell{a011} \eeqa
After implementing the crucial step mentioned above, the closed form of the higher
derivative corrections of two scalars and two gauge fields (which must be added to DBI)\cite{Hatefi:2012ve} turned out
to be
\beqa (2\pi\alpha')^4\frac{1}{ 2
\pi^2}T_p\left(\alpha'\right)^{n+m}\sum_{m,n=0}^{\infty}(\cL_{8}^{nm}+\cL_{9}^{nm}+\cL_{10}^{nm}),\labell{highder678}\eeqa
\beqa &&\cL_{8}^{nm}=-
\Tr\left(\frac{}{}a_{n,m}\cD_{nm}[D_a \phi^i D^b \phi_i F^{ac}F_{bc}]+ b_{n,m}\cD'_{nm}[D_a \phi^i F^{ac} D^b \phi_i F_{bc}]+h.c.\frac{}{}\right),\nonumber\\
&&\cL_{9}^{nm}=-\Tr\left(\frac{}{}a_{n,m}\cD_{nm}[D_a \phi^i D^b \phi_i F_{bc}F^{ac}]+\frac{}{}b_{n,m}\cD'_{nm}[D_a \phi^i F_{bc} D^b \phi_i F^{ac}]+h.c.\frac{}{}\right),\nonumber\\
&&\cL_{10}^{nm}=\frac{1}{2}\Tr\left(\frac{}{}a_{n,m}\cD_{nm}[D_a
\phi^i D^a \phi_i F^{bc}F_{bc}]+\frac{}{}b_{n,m}\cD'_{nm}[D_a \phi^i
F_{bc} D^a \phi_i F^{bc}]+h.c\frac{}{}\right),\nonumber
\eeqa
As usual, the above couplings are valid up to total derivative terms
and terms such as $\prt_a\prt^aFFD\phi D\phi$ that vanish on-shell.

 The examples above suggest that there exists a regularity in the
  higher derivative expansions. One can formulate a prescription based on this regularity.
 In order to find all infinite higher derivative
corrections we must find the S-matrix element of desired amplitudes which are either non-BPS or BPS amplitudes. The
next step is using the relation between Mandelstam variables. In
 other words we must rewrite the amplitudes
 such that all poles can be seen in a clear way.
 The third step is finding leading couplings from tachyonic DBI or DBI action.
 The last step is to express the symmetric trace in terms of the ordinary trace and apply the
  higher derivative corrections $\cD_{nm},\cD'_{nm}$ (as appeared in \reef{esiesi1}) on it.

\vspace{.2in}
Let us apply the prescription to the current case, the higher derivatives
vertices of four-scalar fields.
The first simple
massless scalar pole is reproduced by the non-abelian kinetic terms
of the scalar field \cite{Garousi:2008xp,Hatefi:2010ik},
 \beqa
&&- T_p(2\pi\alpha')^4{\rm STr}
\left(-\frac{1}{4}D_a\phi^iD_b\phi_iD^b\phi^jD^a\phi_j+\frac{1}{8}
(D_a\phi^i D^a\phi_i)^2\right)\,\, \labell{a011}
\eeqa
Applying our prescription, one can easily determine their higher
derivative forms by noting universality property
 that was present in the previous works as follows
\beqa
(2\pi\alpha')^4\frac{1}{4\pi^2}T_p\left(\alpha'\right)^{n+m}\sum_{m,n=0}^{\infty}(\cL_{11}^{nm}+\cL_{12}^{nm}+\cL_{13}^{nm})\labell{lagrang1}\eeqa
\beqa &&\cL_{11}^{nm}=-
\Tr\left(\frac{}{}a_{n,m}\cD_{nm}[D_{a}\phi^i D_{b}\phi_i D^{b}\phi^j D^{a}\phi_j]+\frac{}{} b_{n,m}\cD'_{nm}[D_{a}\phi^i D^{b}\phi^j  D_{b}\phi_i
D^{a}\phi_j ]+h.c.\frac{}{}\right)\nonumber\\
&&\cL_{12}^{nm}=-\Tr\left(\frac{}{}a_{n,m}\cD_{nm}[D_{a}\phi^i D_{b}\phi_i D^{a}\phi^j D^{b}\phi_j]+\frac{}{}b_{n,m}\cD'_{nm}[D_{b}\phi^i D^{b}\phi^j
 D_{a}\phi_i  D^{a}\phi_j  ]+h.c.\frac{}{}\right)\nonumber\\
&&\cL_{13}^{nm}=\Tr\left(\frac{}{}a_{n,m}\cD_{nm}[D_{a}\phi^i D^{a}\phi_i D_{b}\phi^j D^{b}\phi_j]+\frac{}{}b_{n,m}\cD'_{nm}[D_{a}\phi^i D_{b}\phi^j
 D^{a}\phi_i  D^{b}\phi_j ]+h.c\frac{}{}\right)\nonumber\\
  \label{L5679}
\eeqa
{By comparing \reef{L5679} and \reef{highder}, we see that the corresponding  equations in \reef{L5679} and \reef{highder} have different numerical coefficients. We will comment on this in the conclusion.}
 We now turn to verification of (\ref{L5679}).

\subsection{Pole analyses }

We analyze the infinite massless poles of the string amplitude
(\ref{141u}) for the relevant cases (i.e., $n=p+2$ (infinite
massless scalar poles) and $n=p$ case (infinite gauge field poles
)\footnote{The case $n=p+4$ has only contact terms but no poles.}),
and show that the vertices together with appropriately chosen WZ vertices
reproduce them.

\subsubsection{Massless scalar poles for $n=p+2$ case }

Working out the trace and using special expansions in the previous section,
the massless scalar poles of the string amplitude take
\beqa &&16  \pi^{3}\mu_p \frac{\eps^{a_{0}\cdots a_{p}}
H^{i(p+2)}_{a_0\cdots a_{p}}}{(p+1)!(s+t+u)}\Tr(\lam_1\lam_2\lam_3)
\sum_{n,m=0}^{\infty} c_{n,m}\bigg(us\xi_{3}^i\xi_1.\xi_2
[s^{m}u^{n}+s^{n}u^{m}]\nonumber\\&&+ut\xi_{2}^i\xi_1.\xi_3
[t^{m}u^{n}+t^{n}u^{m}]+ts\xi_{1}^i\xi_3.\xi_2
[s^{m}t^{n}+s^{n}t^{m}]\bigg) \label{smatrix123}\eeqa where we have
normalized the amplitude by $(2^{1/2}\pi^{1/2}\mu_p) $.

\begin{center}
\begin{picture}
(600,125)(0,0)
\Line(125,105)(185,70)\Text(150,105)[]{$\phi_{1}$}
\Line(125,35)(185,70) \Text(150,39)[]{$\phi_{3}$}
\Line(125,70)(185,70) \Text(145,79)[]{$\phi_2$}
\Line(185,70)(255,70) \Text(220,88)[]{$\phi$}
\Gluon(255,70)(295,105){4.20}{5}\Text(275,105)[]{$C_{p+1}$}
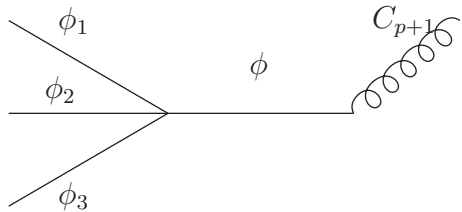
\end{picture}\\ {\sl Figure 2 : The Feynman diagram corresponding to the amplitudes \reef{smatrix123}.}
\end{center}
$\frac{}{}$

$\frac{}{}$
The first massless poles are reproduced by the non-abelian kinetic
terms of the scalar field \reef{a011}. Now we want to show that those higher derivative corrections with
the correct coefficient \reef{lagrang1} reproduce all infinite scalar
poles in the $(s+t+u)$-channel of the string theory S-matrix. Below,
we show that the rest of the poles are reproduced by
\beqa
\lambda\mu_p\int d^{p+1}\sigma {1\over (p+1)!}
(\veps^v)^{a_0\cdots a_{p}}\,\Tr\left(\phi^i\right)\,
H^{(p+2)}_{ia_0\cdots a_{p}}(\sigma)
\eeqa
Let us consider the field theory amplitude of one R-R field and
three scalars for the $p+2=n$ case. In Feynman rules (we use the Feynman
gauge), it is given by
\beqa
{\cal A}&=&V_{\alpha}^{i}(C_{p+1},\phi)G_{\alpha\beta}^{ij}(\phi)V_{\beta}^{j}(\phi,\phi_1,
\phi_2,\phi_3),\labell{amp549}
\eeqa
where
\beqa
G_{\alpha\beta}^{ij}(\phi) &=&\frac{-i\delta_{\alpha\beta}\delta^{ij}}{T_p(2\pi\alpha')^2
k^2}=\frac{-i\delta_{\alpha\beta}\delta^{ij}}{T_p(2\pi\alpha')^2
(t+s+u)},\nonumber\\
V_{\alpha}^{i}(C_{p+1},\phi)&=&i(2\pi\alpha')\mu_p\frac{1}{(p+1)!}(\veps^v)^{a_0\cdots a_{p}}
 H^{i(p+2)}_{a_0\cdots a_{p}}\Tr(\lambda_{\alpha}).
\labell{Fey}
\eeqa
where $k$ is the off-shell momentum of the scalar field and $k^2$ has been replaced
 by $(t+s+u)$ in the propagator in the first equation of (\ref{Fey}). Note that $\Tr(\lambda_{\alpha})$ is just nonzero for the abelian generator $\lambda_{\alpha}$.
Since the off-shell scalar field (i.e., the field $\phi$ in
(\ref{amp549})) is abelian, we must consider 12 (as opposed to
the full 24) cyclic permutations\footnote{ The 12 possible cyclic
permutations of the vertices are associated with different orderings
of generators inside the trace:
\beqa
\Tr(\lam_1\lam_2\lam_3\lambda_{\beta}),
\Tr(\lam_1\lam_2\lambda_{\beta}\lam_3)
\Tr(\lam_1\lambda_{\beta}\lam_2\lam_3),\nonumber\\
\Tr(\lam_2\lam_3\lam_1\lambda_{\beta})
\Tr(\lam_2\lam_3\lambda_{\beta}\lam_1),
\Tr(\lam_2\lambda_{\beta}\lam_3\lam_1)\nonumber\\
\Tr(\lam_3\lambda_{\beta}\lam_1\lam_2),
\Tr(\lam_3\lam_1\lam_2\lambda_{\beta})
\Tr(\lam_3\lam_1\lambda_{\beta}\lam_2),\nonumber\\
\Tr(\lambda_{\beta}\lam_1\lam_2\lam_3)
\Tr(\lambda_{\beta}\lam_3\lam_1\lam_2),
\Tr(\lambda_{\beta}\lam_2\lam_3\lam_1)
\eeqa
for the given 123  ordering of the amplitude.
} of the vertices for the given $\Tr(\lam_1\lam_2\lam_3)$ ordering.
 After including the permutations, one obtains the higher derivative vertex  $ V_{\beta}^{j}(\phi,\phi_1,
\phi_2,\phi_3)$  from the higher derivative couplings
in \reef{lagrang1} as follows:
\vspace{.1in}
\beqa
V_{\beta}^{j}(\phi,\phi_1, \phi_2,\phi_3)
 &=&\Tr(\lam_1\lam_2\lam_3\lambda_{\beta}) I_9 \;[V_{1}^{j}(\phi,\phi_1,
\phi_2,\phi_3)+V_{2}^{j}(\phi,\phi_1,
\phi_2,\phi_3)+V_{3}^{j}(\phi,\phi_1,
\phi_2,\phi_3)]\nonumber\eeqa
where
\beqa
I_9&=&\frac{1}{4\pi^2}(\alpha')^{n+m}(a_{n,m}+b_{n,m})(2\pi\alpha')^4T_{p}
\label{I9def}
\eeqa
and
\beqa
 V_{1}^{j}(\phi,\phi_1,
\phi_2,\phi_3)&=&\frac{ts}{2}\xi_{1}^j\xi_2.\xi_3
\bigg((k_3\inn k_1)^m(k_1\inn k_2)^n+(k_3\inn k)^n(k_1\inn k_3)^m
+(k_1\inn k_2)^n(k\inn k_2)^m\nonumber\\&&+(k\inn k_3)^n (k\inn k_2)^m
+(k\inn k_2)^n(k_2\inn k_1)^m+(k_3\inn k_1)^n(k_2\inn k_1)^m
\nonumber\\&&+(k_2\inn k)^n(k_3\inn k)^m+(k_3\inn k_1)^n(k_3\inn k)^m
\bigg),\nonumber\eeqa
\beqa
 V_{2 }^{j}(\phi,\phi_1,\phi_2,\phi_3)&=&\frac{ut}{2}\xi_{2}^j\xi_1.\xi_3
\bigg((k_2\inn k_1)^m(k_3\inn k_2)^n+(k_2\inn k_1)^m(k_1\inn k)^n
+(k_2\inn k_3)^m(k\inn k_3)^n\nonumber\\&&+(k_1\inn k_2)^n (k_3\inn k_2)^m
+(k_3\inn k_2)^n(k_3\inn k)^m+(k\inn k_1)^n(k_3\inn k)^m
\nonumber\\&&+(k_3\inn k)^n(k_1\inn k)^m+(k_2\inn k_1)^n(k_1\inn k)^m
\bigg),\labell{verpppp}\eeqa
\beqa
 V_{3 }^{j}(\phi,\phi_1,\phi_2,\phi_3)&=&\frac{us}{2}\xi_{3}^j\xi_1.\xi_2
\bigg((k\inn k_1)^m(k_3\inn k_1)^n+(k\inn k_1)^m(k_2\inn k)^n
+(k_1\inn k_3)^m(k\inn k_1)^n\nonumber\\&&+(k_1\inn k_3)^m (k_3\inn k_2)^n
+(k_3\inn k_1)^n(k_3\inn k_2)^m+(k\inn k_2)^n(k_3\inn k_2)^m
\nonumber\\&&+(k_1\inn k)^n(k_2\inn k)^m+(k_2\inn k_3)^n(k_2\inn k)^m
\bigg)\nonumber
\eeqa
The coefficients $a_{n,m}$ and $b_{n,m}$ are identical to those
that were computed in \cite{Hatefi:2010ik} for the case of four-gauge fields
amplitude.
We list some of them for convenience and self-containedness of the paper:
\beqa
&&a_{0,0}=-\frac{\pi^2}{6},\,b_{0,0}=-\frac{\pi^2}{12},a_{1,0}=2\z(3),\,a_{0,1}=0,\,b_{0,1}=-\z(3),a_{1,1}=a_{0,2}=-7\pi^4/90,\nonumber\\
&&a_{2,2}=(-83\pi^6-7560\z(3)^2)/945,b_{2,2}=-(23\pi^6-15120\z(3)^2)/1890,a_{1,3}=-62\pi^6/945,\nonumber\\
&&\,a_{2,0}=-4\pi^4/90,\,b_{1,1}=-\pi^4/180,\,b_{0,2}=-\pi^4/45,a_{0,4}=-31\pi^6/945,a_{4,0}=-16\pi^6/945,\nonumber\\
&&a_{1,2}=a_{2,1}=8\z(5)+4\pi^2\z(3)/3,\,a_{0,3}=0,\,a_{3,0}=8\z(5),b_{1,3}=-(12\pi^6-7560\z(3)^2)/1890,\nonumber\\
&&a_{3,1}=(-52\pi^6-7560\z(3)^2)/945, b_{0,3}=-4\z(5),\,b_{1,2}=-8\z(5)+2\pi^2\z(3)/3,\nonumber\\
&&b_{0,4}=-16\pi^6/1890.\eeqa
Substituting $k_1\inn k=k_2.k_3-(k^2)/2$, $k_3\inn
k=k_2.k_1-(k^2)/2$ and $k_2\inn k=k_1.k_3-(k^2)/2$, one finds the
following massless scalar poles\footnote{ Contact terms are produced as
well when the terms $k^2$ in the vertex \reef{verpppp} get canceled
against the $k^2$ in the denominator of the scalar field propagator.
We will not consider them explicitly (for more details see section
7.2 of \cite{Hatefi:2012wj}) }, \beqa
&&32\pi\mu_p\frac{\eps^{a_{0}\cdots a_{p}} H^{i(p+2)}_{a_0\cdots
a_{p}}}{(p+1)!(s+t+u)}\Tr(\lam_1\lam_2\lam_3)
\sum_{n,m=0}^{\infty}(a_{n,m}+b_{n,m})\bigg(us\xi_{3i}\xi_1.\xi_2
[s^{m}u^{n}+s^{n}u^{m}]\nonumber\\&&+ut\xi_{2i}\xi_1.\xi_3
[t^{m}u^{n}+t^{n}u^{m}]+ts\xi_{1i}\xi_3.\xi_2
[s^{m}t^{n}+s^{n}t^{m}]\bigg) \label{amphigh8} \eeqa
Let us compare this with the massless scalar poles
of the string theory amplitude \reef{smatrix123}. We  have chosen several
 values of $n,m$. Note that for simplicity common factors of both
string and field theory have been omitted. For $n=m=0$, the amplitude \reef{amphigh8} has the following coefficient
\beqa
-2(a_{0,0}+b_{0,0})&=&-2(\frac{-\pi^2}{6}+\frac{-\pi^2}{12})=\frac{\pi^2}{2}\nonumber
\eeqa
The string amplitude has a corresponding term with a numerical factor of $(\pi^2 c_{0,0})$.
 It indeed matches with the field theory result. At the order of $\alpha'$, the field theory result
 \reef{amphigh8} has a term with the coefficient $(a_{1,0}+a_{0,1}+b_{1,0}+b_{0,1})$. This vanishes
 as   the corresponding string theory term vanishes with the coefficent $\pi^2 (c_{1,0}+c_{0,1})$.
 At the  order of $(\alpha')^2$, the amplitude \reef{amphigh8} has the following coefficient
\beqa
-2(a_{0,2}+a_{2,0}+b_{0,2}+b_{2,0})\bigg(us\xi_{3i}\xi_1.\xi_2 [s^2+u^2]+ut\xi_{2i}\xi_1.\xi_3 [t^2+u^2]+ts\xi_{1i}\xi_3.\xi_2 [s^2+t^2]\bigg)\nonumber\\
-2(a_{1,1}+b_{1,1})\bigg(us\xi_{3i}\xi_1.\xi_2 [2su]+ut\xi_{2i}\xi_1.\xi_3 [2tu]+ts\xi_{1i}\xi_3.\xi_2 [2st]\bigg)\nonumber\\
=\frac{\pi^4}{3}\bigg(us\xi_{3i}\xi_1.\xi_2
[s^2+u^2]+ut\xi_{2i}\xi_1.\xi_3 [t^2+u^2]+ts\xi_{1i}\xi_3.\xi_2
[s^2+t^2]\bigg)\nonumber\\+\frac{\pi^4}{6}\bigg(us\xi_{3i}\xi_1.\xi_2
[2su]+ut\xi_{2i}\xi_1.\xi_3 [2tu]+ts\xi_{1i}\xi_3.\xi_2 [2st]\bigg)
\nonumber \eeqa
and the string result has
 \beqa
 \pi^2(c_{2,0}+c_{0,2})\bigg(us\xi_{3i}\xi_1.\xi_2 [s^2+u^2]+ut\xi_{2i}\xi_1.\xi_3 [t^2+u^2]+ts\xi_{1i}\xi_3.\xi_2 [s^2+t^2]\bigg)\nonumber\\
 +\pi^2 c_{1,1}\bigg(us\xi_{3i}\xi_1.\xi_2 [2su]+ut\xi_{2i}\xi_1.\xi_3 [2tu]+ts\xi_{1i}\xi_3.\xi_2 [2st]\bigg)
 \nonumber
 \eeqa
The latter becomes equal to the former upon using the coefficients
in \reef{hash}. At the order of $\alpha'^3$, field theory
amplitude has two different terms. The first one has the coefficient
of $(a_{3,0}+a_{0,3}+b_{0,3}+b_{3,0})$ which is zero and the
corresponding term on the string theory side has a
$\pi^2(c_{0,3}+c_{3,0})$ coefficient which is again zero in
accordance with field theory. The second term has the following
coefficient
\beqa
&&-2(a_{1,2}+a_{2,1}+b_{1,2}+b_{2,1})=-8\pi^2\z(3)\nonumber\eeqa
and it again matches with the corresponding coefficient in the
string amplitude, $\pi^2 (c_{2,1}+c_{1,2})=-8\pi^2\z(3)$.

The other comparisons to all orders of $\alpha'$ need not be done\footnote{The same checks were carried out for finding infinite massless poles of $<V_CV_AV_AV_A>$
and $<V_CV_\phi V_AV_A>$in \cite{Hatefi:2010ik} and
\cite{Hatefi:2012ve}. }. Therefore we could exactly reproduce
the
infinite massless scalar poles of the string theory amplitude of
  one RR and three scalar fields in the worldvolume of BPS branes. This confirms that we have obtained the
higher derivative couplings of four scalars with the correct
coefficients and they are exact up to terms that vanish on-shell.

\subsubsection{Massless gauge field poles for $p=n$ case }

 Working out the trace, it is possible to obtain all massless gauge poles in the string theory side as follows:
\beqa
{\cal A}_{2}&=&\pm 2^{-1/2}(2^{1/2}\pi^{1/2}\mu_p)\frac{32}{2p!}L_2\bigg\{
 \eps^{a_{0}\cdots a_{p-2}ba}H^{k (p)}_{a_{0}\cdots a_{p-2}}\bigg(2us\xi_{1}.\xi_{2}k_{2a}k_{3b}\xi_{3k}
+2ut\xi_{3}.\xi_{1}k_{2a}k_{3b}\xi_{2k}\nonumber\\&&+2st\xi_{3}.\xi_{2}k_{2a}k_{3b}\xi_{1k}\bigg)
+p^k \eps^{a_{0}\cdots a_{p-1}a}H^{(p)}_{a_{0}\cdots a_{p-1}}\bigg(-2usk_{2a}\xi_1.\xi_2\xi_{3k}
+2ut k_{3a}\xi_1.\xi_3\xi_{2k}\bigg)\bigg\}\nonumber
\eeqa
 where the amplitude is normalized by $2^{1/2}\pi^{1/2}\mu_{p}$.
Substituting special expansion of (\ref{expansion44}) into the amplitude and keeping
all the gauge field poles (but not the contact terms), one gets
\beqa
{\cal A}_{2}&=&\pm  \frac{32}{2p!} \pi^{2}\mu_p \bigg\{
   \sum_{n=-1}^{\infty}\frac{1}{u}{b_n(t+s)^{n+1}}(2\xi_{2}.\xi_{3}k_{2a}k_{3b}\xi_{1k})\eps^{a_{0}\cdots a_{p-2}ba}H^{k(p)}_{a_{0}\cdots a_{p-2}} \nonumber\\&&+\bigg(\bigg[\sum_{n=-1}^{\infty}\frac{1}{t}{b_n(u+s)^{n+1}}\bigg(\eps^{a_{0}\cdots a_{p-2}ba}H^{k(p)}_{a_{0}\cdots a_{p-2}} (2\xi_{1}.\xi_{2}k_{2a}k_{3b}\xi_{3k})+p^k \eps^{a_{0}\cdots a_{p-1}a}H^{(p)}_{a_{0}\cdots a_{p-1}}\nonumber\\&&\times (-2k_{2a}\xi_1.\xi_2\xi_{3k})\bigg)\bigg]
  -\bigg[2\leftrightarrow 3\bigg]\bigg)\bigg\}\Tr(\lam_1\lam_2\lam_3)\label{esi56}
\eeqa
Let us examine the massless gauge poles. First, we will show that effective
field theory reproduces the infinite massless gauge poles in the
$u$-channel, i.e., the first term in \reef{esi56}. Then we reproduce
the second and third terms in \reef{esi56}. Since the amplitudes
in $t$- and $s$-channels are similar, we just reproduce all infinite massless gauge t-channel poles in detail. By interchanging the momentum and
polarization of scalar fields , $(2\leftrightarrow 3)$, one can find
the other infinite massless gauge poles in the s-channel as well. The needed field
theory vertex for the first term in \reef{esi56} is
\beqa S^{(5)}&=&i\lambda\mu_p\int \STr\left(F P\left[
C^{(p-1)}(\s,\phi)\right]\right)
\nonumber\\
&=&{i}\lambda^2\mu_p\int d^{p+1}\s {1\over (p-1)!}(\veps^v)^{a_0\cdots a_{p}}
\left[\Tr\left(F_{a_{0}a_{1}}\partial_{a_{2}}\phi^k\right)
 C^{(p-1)}_{k a_{3}... a_{p}}(\s)\right].
\label{fin1}
\eeqa
Where the scalar field comes from pull-back (see section 5 of
\cite{Hatefi:2012wj}). The partial derivatives on the scalar fields
can be replaced by the covariant derivatives. The connection
  parts do not contribute because there is no external gauge field; therefore, the off-shell gauge
  field must come from the abelian field strength.
 With this vertex, the massless gauge poles in the $u$-channel are reproduced in the form
\beqa
{\cal A}&=&V^a_{\alpha}(C_{p-1},\phi_1, A)G^{ab}_{\alpha\beta}(A)V^b_{\beta}(A,\phi_2,\phi_3),\labell{amp642}
\eeqa
where the vertices and gauge field propagator are
\beqa
V^a_{\alpha}(C_{p-1},\phi_1,A)&=&\frac{\lambda^2\mu_p}{(p)!}(\veps^v)^{a_0\cdots a_{p-1}a}(H^{(p)})^k{}_{a_0\cdots a_{p-2}}\xi_{1k}k_{a_{p-1}}\Tr(\lam_1\lambda_\alpha)\sum_{n=-1}^{\infty}b_n(\alpha'k_1.k)^{n+1},\nonumber\\
\eeqa
with
\beqa
V_\beta^{b}(A,\phi_2,\phi_3)&=&iT_p(2\pi\alpha')^{2} \xi_{2}.\xi_{3}(k_2-k_3)^{b}\Tr(\lambda_{2}\lambda_{3}\lambda_{\beta}),
\nonumber\\
G_{\alpha\beta}^{ab}(A)&=&\frac{i\delta_{\alpha\beta}\delta^{ab}}{(2\pi\alpha')^{2}T_p(k^2)},
\nonumber
\eeqa
$k$ is the momentum of the abelian gauge field, $k^2=(k_3+k_2)^2=-u$.
The propagator is found from the kinetic term of the gauge field in the Born-Infeld action.
The vertex of $V_\beta^{b}(A,\phi_2,\phi_3)$ has been obtained from the kinetic term of the scalar field $\frac{\lambda^2}{2}  \Tr(D_a\phi_i D^a\phi^i)$. (A vertex similar to $V_\beta^{b}(A,\phi_2,\phi_3)$ was obtained in \cite{Hatefi:2012wj}.)
The simple massless poles of string amplitude indicate that the kinetic term of the scalar field
has no higher derivative corrections; hence, the vertex
$V_\beta^{b}(A,\phi_2,\phi_3)$ has no higher derivative correction
either.
 The vertex $V^a_{\alpha}(C_{p-1},\phi_1,A)$ must be derived from the higher derivative extensions of the WZ coupling \reef{fin1} as follows
\beqa S^{(5)}&=&{i}\lambda^2\mu_p\int d^{p+1}\s {1\over
(p-1)!}(\veps^v)^{a_0\cdots a_{p}} \nonumber \\&&\times
\sum_{n=-1}^{\infty}
b_n(\alpha')^{n+1}\left[\Tr\bigg(\partial_{a_{m_{0}}}\cdots
\partial_{a_{m_{n}}} F_{a_{0}a_{1}}\partial^{a_{m_{0}}}\cdots
\partial^{a_{m_{n}}}\partial_{a_{2}}\phi^k\bigg)
  C^{(p-1)}_{k a_{3}... a_{p}}(\s)\right]
\label{fin167}
\eeqa
Inserting this into the amplitude \reef{amp642}, one finds
\beqa
{\cal A}&=&(2\pi\alpha')^{2}\frac{\mu_p}{p!u}\eps^{a_{0}\cdots a_{p-1}a}H^{(p)}_{ka_{0}\cdots a_{p-2}}\xi^{k}_{1}
\Tr(\lambda_{1}\lambda_{2}\lambda_{3})\sum_{n=-1}^{\infty}b_n\bigg(\frac{\alpha'}{2}\bigg)^{n+1}(s+t)^{n+1}
\nonumber\\&&\times\bigg(-2(\xi_2.\xi_3)k_{2a} k_{3_{a_{p-1}}}\bigg).\labell{ver22}
\eeqa
These are precisely the $u$-channel massless poles of \reef{esi56}. Unlike the $p+2=n$ case in the
previous section, here there are no residual contact terms.
\begin{center}
\begin{picture}
(600,105)(0,0)
\Line(150,105)(200,70)\Text(175,105)[]{$\phi_{2}$}
\Line(150,35)(200,70)\Text(175,39)[]{$\phi_{3}$}
\Photon(200,70)(250,70){4}{7.5}\Text(225,90)[]{$A$}
\Gluon(250,70)(300,105){4.20}{5}\Text(268,100)[]{$C_{p-1}$}
\Line(250,70)(300,35)\Text(275,39)[]{$\phi_{1}$}
\SetColor{Black}
\Vertex(200,70){1.5} \Vertex(250,70){1.5}
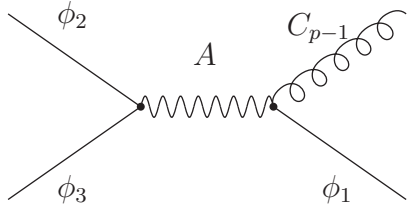
\end{picture}\\ {\sl Figure 3 : The Feynman diagram  corresponding to the massless gauge pole of the amplitude \reef{esi56} in u-channel.}
\end{center}

Having reproduced all infinite poles corresponding to the first term of \reef{esi56}, we turn to the rest
of the terms (namely, we would like to reproduce the second and third terms of \reef{esi56}). We quote
those terms here:
\beqa
{\cal A}&=&\pm  \frac{32}{2p!} \pi^{2}\mu_p \bigg\{
   \sum_{n=-1}^{\infty}\frac{1}{t}{b_n(u+s)^{n+1}}\bigg(\eps^{a_{0}\cdots a_{p-2}ba}H^{k(p)}_{a_{0}\cdots a_{p-2}} (2\xi_{1}.\xi_{2}k_{2a}k_{3b}\xi_{3k})+p^k \eps^{a_{0}\cdots a_{p-1}a}\nonumber\\&&\times H^{(p)}_{a_{0}\cdots a_{p-1}}(-2k_{2a}\xi_1.\xi_2\xi_{3k})\bigg)\bigg\}\Tr(\lam_1\lam_2\lam_3)\label{esi57}
\eeqa
Again it vanishes for the abelian group. We just kept the infinite massless poles in the t-channel and at the moment we disregard all contact terms. We now show that effective field theory will result in the infinite massless gauge poles in the t-channel\footnote{ By interchanging $(2\leftrightarrow 3)$ we find the other massless poles in the s-channel.}. The corresponding effective field theory vertex is given by
\beqa S^{(6)}&=&i\lambda\mu_p\int \STr\left(F P\left[
C^{(p-1)}(\s,\phi)\right]\right)
\nonumber\\
&=&{i}\lambda^2\mu_p\int d^{p+1}\s {1\over (p-1)!}(\veps^v)^{a_0\cdots a_{p}}
\left[\Tr\left(F_{a_{0}a_{1}}\phi^k\right)
 \partial_{k}C^{(p-1)}_{ a_{2}... a_{p}}(\s)\right]\,\,\,
\label{fin178}
\eeqa
Notice that the scalar field in \reef{fin178} comes from Taylor expansion (for review, see
section 5 of \cite{Hatefi:2012wj}). Now if we extract the field
strength and take integration by parts we will have several terms such as the following:
\beqa
S^{(6)} &=& i\lambda^2\mu_p\int d^{p+1}\s
{1\over (p-1)!}(\veps^v)^{a_0\cdots a_{p}}
\bigg(-A_{a_{1}}\partial_{a_{0}}\phi^k
 \partial_{k}C^{(p-1)}_{ a_{2}... a_{p}}(\s)-A_{a_{1}}\phi^k \partial_{a_{0}} \partial_{k}C^{(p-1)}_{ a_{2}... a_{p}}(\s)\bigg)
\nonumber
\eeqa

\vspace{.1in}

Having taken into account the off-shell gauge field, writing the above coupling in momentum space and applying momentum
conservation along the world volume of brane we can obtain the final
form of the vertex of one off-shell gauge field, one RR $p-1$ form
field and one external scalar field (which we labeled its polarization with
$\xi_3$). The vertex $V^a_{\alpha}(C_{p-1},\phi_3,A)$ should be obtained from
the higher derivative extension of the WZ coupling \reef{fin178}
 as
\beqa S^{(6)}&=&{i}\lambda^2\mu_p\int d^{p+1}\s {1\over
(p-1)!}(\veps^v)^{a_0\cdots a_{p}} \nonumber \\&&\times
\sum_{n=-1}^{\infty}
b_n(\alpha')^{n+1}\left[\Tr\bigg(\partial_{a_{m_{0}}}\cdots
\partial_{a_{m_{n}}} F_{a_{0}a_{1}}\partial^{a_{m_{0}}}\cdots
\partial^{a_{m_{n}}} \phi^k\bigg)
  \partial_{k}C^{(p-1)}_{ a_{2}... a_{p}}(\s)\right]\,\,\,
\label{fin1698}
\eeqa
Therefore
 \beqa
V^a_{\alpha}(C_{p-1},\phi_3,A)&=&\frac{\lambda^2\mu_p}{(p-1)!}(\veps^v)^{a_0\cdots
a_{p-1}a}\Tr(\lam_3\lambda_\alpha)\sum_{n=-1}^{\infty}b_n(\alpha'k_3.k)^{n+1}
\nonumber\\&&\times  p^k \xi_{3k}(p+k_3)_{a_{p-1}}(C^{(p-1)})_{a_0\cdots a_{p-2}},
\eeqa
Note that the above vertex is taken into account after applying
higher derivative extensions in \reef{fin1698}. The corresponding
Feynman amplitude is now
\beqa {\cal
A}&=&V^a_{\alpha}(C_{p-1},\phi_3,
A)G^{ab}_{\alpha\beta}(A)V^b_{\beta}(A,\phi_1,\phi_2),\labell{amp6429}
\eeqa
with
\beqa
V_\beta^{b}(A,\phi_1,\phi_2)&=&iT_p(2\pi\alpha')^{2}
\xi_{1}.\xi_{2}(k_1-k_2)^{b}\Tr(\lambda_{1}\lambda_{2}\lambda_{\beta}),
\nonumber\\
G_{\alpha\beta}^{ab}(A)&=&\frac{i\delta_{\alpha\beta}\delta^{ab}}{(2\pi\alpha')^{2}T_p(k^2)},\nonumber\eeqa
$k^2$ in the propagator is now, $k^2=(k_1+k_2)^2=-t$. Now applying momentum conservation we can reexpress $V_\beta^{b}(A,\phi_1,\phi_2)$ as:
\beqa
V_\beta^{b}(A,\phi_1,\phi_2)&=&iT_p(2\pi\alpha')^{2} \xi_{1}.\xi_{2}(-2k_2-k_3-p)^{b}\Tr(\lambda_{1}\lambda_{2}\lambda_{\beta})
\nonumber
\eeqa
Replacing them in the amplitude \reef{amp6429}, we find that the infinite massless gauge poles in the t-channel are reproduced as
 \beqa
{\cal A}&=&(2\pi\alpha')^{2}\frac{\mu_p}{(p)!t}\eps^{a_{0}\cdots a_{p-1}b}\xi_{3k}
(\xi_2.\xi_1)\Tr(\lambda_{1}\lambda_{2}\lambda_{3})\sum_{n=-1}^{\infty}b_n\bigg(\frac{\alpha'}{2}\bigg)^{n+1}(s+u)^{n+1}
\nonumber\\&&\times \bigg(-2 k_{2b} p^k H^{(p)}_{ a_{0}... a_{p-1}}+2 k_{2b} k_{3a_{p-1}}  H^{k(p)}_{ a_{0}... a_{p-2}}\bigg).\labell{ver22final11}
\eeqa
These exactly are t-channel massless poles of the string theory
amplitude \reef{esi57}. So we find that field theory computations are in exact agreement with the string amplitude at pole
levels.  Similar computations in $s$-channel  also lead to
agreement.


The simple massless poles of string amplitude show that the
kinetic term of the scalar fields
has no higher derivative corrections so the vertex $V_\beta^{b}(A,\phi_1,\phi_2)$ has no higher derivative correction either.

\vskip 0.1in

\subsection{Contact term analyses}

Above we have successfully determined the field theory amplitudes
that reproduce all of the poles of the string amplitude. We attempt
to achieve the same for the contact terms below. We have succeeded
in the case of $p+4=n$. However, as for the cases of $n=p+2, n=p$,
we could not find the field theory vertices that reproduce
the leading order contact terms nor the infinite extension.
{Perhaps this is a hint that the pull-back method may need modification.}

\subsubsection{Contact terms for $p+4=n$ case}

The relevant part of the string amplitude can be rewritten as
\beqa {\cal A}_{2}&=&\pi^{2}\mu_p
\xi_{1i}\xi_{2j}\xi_{3k}\Tr(P_{-}\fsH_{(n)}M_p
\Gamma^{kji})\nonumber\\
&&\times\bigg(\sum_{n=-1}^{\infty}b_n(u+s)^{n+1}-\sum_{p,n,m=0}^{\infty}e_{p,n,m}t^{p+1}(su)^{n}(s+u)^m\bigg)
\labell{cterms1} \eeqa

\begin{picture}
(600,120)(0,0)
\Line(195,105)(245,70) \Text(190,95)[]{$\phi_2$}
\Line(195,35)(245,70) \Text(190,52)[]{$\phi_3$}
\Gluon(245,70)(295,105){4.20}{5}\Text(264,100)[]{$C_{p+3}$}
\Line(245,70)(295,35) \Text(260,50)[]{$\phi_1$}
\end{picture}\\ {\sl Figure 4 : The Feynman diagram corresponding to the amplitude of \reef{cterms1}.}

$\frac{}{}$

$\frac{}{}$

The contact terms of this amplitude can be reproduced by an infinite
extension of a Wess-Zumino term. Let us reproduce all those contact
terms using a vertex that contains
 a ($p$+3)-form Ramond-Ramond
potential and three scalar fields. There are two relevant
Wess-Zumino terms depending on whether the third scalar comes from
Taylor expansion or pull-back. The first contribution is given by
 \beqa
S_7&=&{1\over2}\lambda^2\mu_p\int d^{p+1}\s {1\over
(p+1)!}(\veps^v)^{a_0\cdots a_{p}}
\Tr\left([\Phi^i,\Phi^j]\Phi^l\right) \prt_l C^{(p+3)}_{jia_0\cdots
a_{p}}(\sigma)\,\,\, . \nonumber \eeqa
The commutator comes from the exponential in Wess-Zumino action and
the last scalar comes from Taylor expansion.
 The second Wess-Zumino term is given by
 \beqa
S_8 &=&{1\over2}\lambda^2\mu_p\int d^{p+1}\s {1\over
(p)!}(\veps^v)^{a_0\cdots a_{p}} \vphantom{1\over p+1}
\Tr\left([\Phi^i,\Phi^j]\,\prt_{a_{0}}\!\Phi^l\right)
C^{(p+3)}_{jila_1\cdots a_{p}}(\sigma) . \nonumber \eeqa
where the commutator comes from the exponential in Wess-Zumino
action and the third scalar comes from pull-back. Applying
integration by parts these two results can be combined to yield the
final result,
\beq S_{9}={1\over3}\lambda^2\mu_p\int d^{p+1}\s {1\over (p+1)!}
(\veps^v)^{a_0\cdots a_{p}}\,\Tr\left(\Phi^i\Phi^j\Phi^l\right)
H^{(p+4)}_{ijla_0\cdots a_{p}}(\sigma)\,\,\,, \labell{final
contactterms} \eeq
where $S_{9}=S_7+S_8$ and $H^{(p+4)}=dC^{(p+3)}$. While the leading
contact terms of the string amplitude $<V_C V_\phi V_\phi V_\phi>$
 are reproduced by \reef{final contactterms}, the rest of the contact terms require
 a higher derivative extension thereof. Evaluating the trace in \reef{cterms1}, one can show that the first term in  \reef{cterms1} can be reproduced by
\beqa {\lambda^2\mu_p \over 3(p+1)!}\int d^{p+1}\s
(\veps^v)^{a_0\cdots a_{p}}\,\sum_{n=-1}^{\infty}b_n
(\alpha')^{n+1}\Tr\left(D^{a_{0}}\cdots
D^{a_{n}}(\Phi^i\Phi^j)D_{a_{0}}\cdots D_{a_{n}}\Phi^k\right)
H^{(p+4)}_{ijka_0\cdots a_{p}} \nonumber
\eeqa
 Since we are looking for interaction between one RR and three scalars, the covariant derivatives of scalar fields can be replaced by their partial derivatives. All of the infinite contact terms in the second term of \reef{cterms1} can be reproduced by
\beqa -{\lambda^2\mu_p \over 3(p+1)!}\int d^{p+1}\s
(\veps^v)^{a_0\cdots a_{p}}\,
\sum_{p,n,m=0}^{\infty}e_{p,n,m}(\frac{\alpha'}{2})^{p+1}
(\alpha')^{2n+m}H^{(p+4)}_{ijka_0\cdots a_{p}}\Tr\bigg((D^a
D_a)^{p+1} D^{a_{1}}... D^{a_{m}}\nonumber\\\times (D^{a_{1}}\cdots
D^{a_{n}}\Phi^i D^{a_{n+1}}\cdots D^{a_{2n}}\Phi^j)D_{a_{1}}\cdots
D_{a_{2n}}D_{a_{1}}\cdots D_{a_{m}}\Phi^k \bigg) \nonumber
\eeqa
\section{Conclusion}

In this work, we have analyzed the amplitudes of
$<V_{C}V_{\phi}>$, $<V_{C}V_{A}>$, $<V_{C}V_{\phi}V_{\phi}>$ and $<V_{C}V_{\phi}V_{\phi}V_{\phi}>$.
 We have found the field theory vertices that reproduce all infinite contact
terms of two- and three- point amplitudes. We could produce all poles of the four-point function,
but we could produce the contact terms only for $p+4=n$ case.
At the moment it is not entirely clear how to produce all infinite contact terms of
$<V_{C}V_{\phi}V_{\phi}V_{\phi}>$ for $p+2=n,\;p=n$ cases. Possibly the pull-back method
may need modification.

\vspace{.2in}
We found universality in all order $\alpha'$ higher derivative corrections of non-BPS and BPS branes and
the universality played an important role in the determination of field
theory vertices.  Several remarks on T-duality are in order.
T-duality can be straightforwardly employed to deduce a pure open
string tree amplitude of scalar vertex operators from a tree
amplitude of gauge field vertex operators. Once one considers an amplitude
of a mixture of open and closed strings, direct computation is necessary because
of the subtleties associated with T-duality. Two subtleties exist in the very construction of
the RR $C$ vertex operator in (\ref{d4Vs3}).
First, the construction of
the $C$ vertex operator was such that one set of oscillators was
used instead of two.\footnote{Perhaps this step may be some kind of analytic continuation.}
The second issue - which was addressed in footnote 4 of \cite{Park:2008sg} - is that the $C$ vertex operator does not contain winding modes, and this must be related
to the fact that we have pointed out above (\ref{point}): the terms that contain
$p^i$ are absent in $<V_C V_A V_A V_A>$.

\vspace{.1in}

We hope to be able to compute higher point amplitudes of various
mixtures of open string and closed string states. Another more
ambitious direction would be to make progress in the full form of
the DBI action. We hope to report on these issues in the near
future.

\vspace{.3in}  
\section*{Acknowledgments}
E.H would like to thank K.S.Narain, F.Quevedo, Joe Polchinski, R.C Myers, M.Douglas  and L.Alvarez-Gaume  for helpful conversations.

\newpage

\end{document}